\documentclass[12pt,onecolumn,draftclsnofoot]{IEEEtran}

\usepackage{hyperref}
\usepackage{amsmath}
\usepackage{amsfonts}
\usepackage{soul}
\usepackage{amsmath}

\usepackage{cite}
\usepackage{amsmath,amssymb,amsfonts}
\usepackage[noend]{algpseudocode}
\usepackage{flushend}

\usepackage{graphicx}
\usepackage{textcomp}
\usepackage{gensymb}
\usepackage{xcolor}
\usepackage{tabularx}
\usepackage{color}
\usepackage{xfrac}
\usepackage{epstopdf}
\AppendGraphicsExtensions{.tif}
\usepackage{booktabs}
\usepackage{paralist}
\usepackage{enumitem}
\usepackage[export]{adjustbox}
\usepackage{multicol}
\usepackage{gensymb}
\usepackage{mathtools}
\usepackage{subcaption}
\usepackage{wrapfig}
\usepackage{circuitikz}
\usepackage{makecell}
\usepackage[normalem]{ulem}
\usepackage{soul}
\usepackage{comment}
\usepackage{fancyhdr}
\usepackage{algpseudocode}
\usepackage[linesnumbered,ruled,vlined]{algorithm2e}
\usepackage{multirow}
\DeclareUnicodeCharacter{2212}{-}
\def\BibTeX{{\rm B\kern-.05em{\sc i\kern-.025em b}\kern-.08em
    T\kern-.1667em\lower.7ex\hbox{E}\kern-.125emX}}

\usepackage{cite}

\ifCLASSINFOpdf
  
\else
  
\fi

\newtheorem{definition}{Definition}
\newtheorem{remark}{Remark}
\newtheorem{theorem}{Theorem}
\newtheorem{lemma}{Lemma}
\newtheorem{example}{Example}
\newtheorem{construction}{Construction}
\newtheorem{corollary}{Corollary}

\begin{document}

\title{Secret Sharing for DNA Probability Vectors\\

\author{
		\IEEEauthorblockN{Wenkai Zhang,  Zhiying Wang}\\
		\IEEEauthorblockA{Center for Pervasive Communications and Computing (CPCC) \\ University of California, Irvine, USA
			\\ \{wenkaiz1, zhiying\}@uci.edu
		} 
	}
}

\maketitle
\begin{abstract}

Emerging DNA storage technologies use composite DNA letters, where information is represented by a probability vector, leading to higher information density and lower synthesis costs.
However, it faces the problem of information leakage in sharing the DNA vessels among untrusted vendors. This paper introduces an asymptotic ramp secret sharing scheme (ARSSS) for secret information storage using composite DNA letters. This innovative scheme, inspired by secret sharing methods over finite fields and enhanced with a modified matrix-vector multiplication operation for probability vectors, achieves asymptotic information-theoretic data security for a large alphabet size. 
Moreover, this scheme reduces the number of reading operations for DNA samples compared  to traditional schemes, and therefore lowers the complexity and the cost of DNA-based secret sharing. 
We further explore the construction of the scheme, starting with a proof of the existence of a suitable generator, followed by practical examples. Finally, we demonstrate efficient constructions to support large information sizes, which utilize multiple vessels for each secret share rather than a single vessel.

\end{abstract}

\section{Introduction}
With the vast demand for large-scale information storage, DNA-based data storage is considered as one of the promising candidates of storage technologies. It is particularly appealing due to its high information density and long durability. 
Storing digital information in DNA typically involves encoding data using the DNA nucleotide alphabet $\{A, C, G, T\}$, creating synthetic DNA sequences, and storing them in the vessels. Afterwards, the DNA samples are sequenced and decoded to retrieve the original information \cite{DNA1,DNA2,DNA3,DNA4,DNA5,DNA6,DNA7}. 

In the original DNA storage framework, data is represented by the DNA alphabet of size $4$, which is capable of storing at most $2$ bits of data per nucleotide location. 
One bottleneck of DNA storage systems is the high synthesis cost.
In addressing this issue, a novel DNA storage system using the probabilities of nucleotides in each location, termed composite DNA letters, is introduced \cite{Anavy_DNA_letters, choi, molecular, data, yan2023, wenkai_full, sabary2024survey}.
The idea is to synthesize multiple DNA sequences simultaneously with the desired probabilities. The alphabet size per location is enlarged since the number of probabilities is potentially infinite, and hence the synthesis cost per data bit can be significantly reduced. The channel of composite DNA letters has also been studied from a coding theory perspective in several works \cite{icc2022, kobovich2023m, zhang2024codes, walter2024coding, cohen2024optimizing, sabary2024error, kobovich2025deepdive}. 

\begin{figure}
    \centering
\includegraphics[scale=1.2]{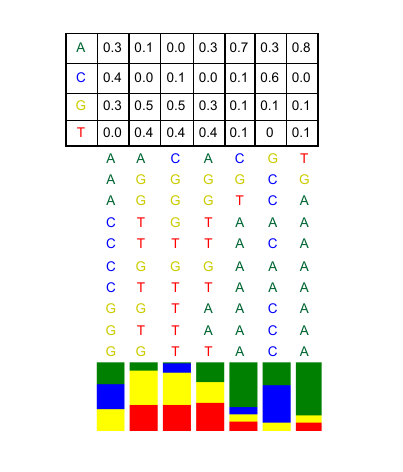}
    \caption{A composite DNA sequence}
    \label{fig: composite}
\end{figure}

A composite DNA letter represents a specific position within a sequence by the predetermined (scaled) \emph{probability vector} $ (x_A, x_C, x_G, x_T)$. Here, each component $x_A, x_C, x_G, x_T$ is a non-negative integer, and their sum $q = x_A + x_C + x_G + x_T$ is a constant parameter called the \emph{resolution}. For instance, $ ((3, 4, 3, 0),(0,1,5,4),\dots)$ characterizes the positions within a composite DNA sequence of resolution $q = 10$. In this case,  there is a probability of $30\%$ for observing A or G, $40\%$ for observing C, and no chance for observing T in the first position; and there is a probability of $10\%$ for observing  C, $50\%$ for G, $40\%$ for T, and no chance for A in the second position, shown in Figure \ref{fig: composite}. In each synthesis cycle, one composite DNA letter (position) is produced by the associated hardware \cite{Anavy_DNA_letters}.

Beyond composite DNA letters, it is worth noting that the exploration of information representation using probabilities has also been studied in molecular communication \cite{pierobon, Farsad, nakano2013molecular}. 
In general, we denote a probability vector as $X=(x_{1},x_{2},\dots,x_{m})$, for some integer $m$. Composite DNA has $m=4$ probability elements, and composite DNA with only two types of nucleotides (e.g., A, C) or particle $Ca^{2+}, Na^{+}$  concentration in molecular communication has $m=2$ probability elements. When $m=2$, we simply write $x=x_1$, because $x_2=q-x_1$ is determined by the first probability, and the probability vector simplifies to a \emph{scalar probability}.

DNA-based storage introduces security concerns in many scenarios \cite{secover,secover1}.
A common issue arises when the storage of DNA vessels is outsourced to untrusted organizations. As DNA-based archiving becomes commercially viable, with many companies joining the revolution \cite{WinNT}, safeguarding sensitive information becomes crucial. Consider a scenario where sensitive secret information needs to be distributed across several untrusted organizations, with each holding a \emph{share} of the secret. The information should only be retrievable when specific combinations of these organizations collaborate, and inaccessible to any other combination. This concept, known as secret sharing, was pioneered by Blakley \cite{sec0} and Shamir \cite{sec2}.

Several papers have focused on preventing unauthorized data access through various encryption methods, such as secret writing \cite{de2016new}, safeguarding against data leakage from sample bleeding \cite{ney2017computer}, DNA-sequence-based encryption, and DNA-structure-based encryption \cite{zhang2022preservation}, among others \cite{DNA_secret0,DNA_secret1,DNA_secret2,2010grammar}). Secret sharing scheme for DNA strands was addressed by Adhikari \cite{Adhikari}, which suffers from low density since each bit of encoded information requires the storage of one DNA sequence.
 
Secret sharing was generalized to \emph{ramp secret sharing schemes} (RSSSs) \cite{ramp1,ramp2,ramp3,ramp4} in order to reduce the sizes of shares by allowing partial information leakage from unauthorized sets of shares. 
In particular, a $(k, L, n)$ RSSS encodes $L$ information shares into $n$ secret shares, ensuring that any $k$ shares determine the secret, and any $k-L+j$ shares leak $\frac{L-j}{L}$ fraction of the secret, for $j = 0,1,\dots,L-1$ (also called weak secrecy). Yamamoto \cite{yamamoto} provided explicit RSSSs both for weak and strong secrecy, where strong secrecy requires that any $L-j$ secret symbols are still protected when $k-L+j$ shares are available, for $j=0,1,\dots,L-1$. 
RSSSs can also be constructed based on explicit MDS codes \cite{MDS, pieprzyk}. RSSSs based on nested codes were proposed in \cite{chen2007} which focuses on secure multiplication computation. Schemes using algebraic geometry codes \cite{Umberto} can achieve efficient communication for ramp secret sharing. 

The ramp secret sharing problem is closely related to the wire-tap channel II described in \cite{wiretapII}. In this model, a legitimate party communicates over a noiseless channel, while an eavesdropper can perfectly access a fixed fraction of the transmitted bits of their choosing. According to Theorem 1 in \cite{wiretapII}, similar principles apply as in ramp secret sharing: Assume $K$ represents the information bits and $N$ the total transmitted bits. If the number of revealed bits is less than or equal to $N−K$, the eavesdropper gains no information; if it is greater than $N−K$, partial information is leaked; and full information is compromised when all $N$ bits are revealed. 
Methods such as random binning, combinatorial argument, coset codes and other coding strategies \cite{wiretapII, V.K., LDPCwire, Errorwire} can achieve the secrecy capacity in wiretap channel II. For finite fields, an information symbol can be kept fully secure by one-time pad, i.e., adding a uniformly random symbol. As will be seen, cost-effective arithmetic operations on composite DNA letters proposed in this paper do not satisfy such a full secret property, hence methods of write-tap channel II require a reduced code rate and a high coding complexity.

\begin{figure}[h]
    \centering
    \includegraphics[scale=0.7]{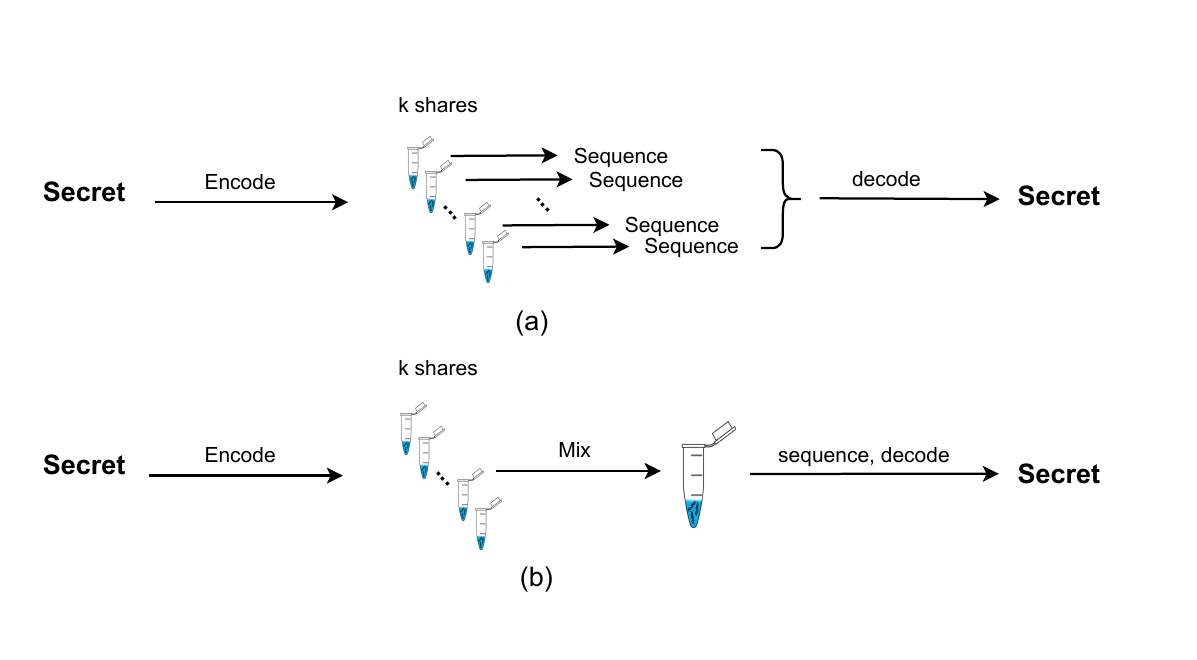}
    \caption{(a) Naive secret sharing scheme; (b) proposed secret sharing scheme. In both cases, $L=1$.}
    \label{fig:scheme}
\end{figure}

Ramp secret sharing for composite DNA letters remains an open problem because of its unique alphabet. 
In particular, the elements in a DNA composite letter are non-negative real numbers whose sum is $q$.  This characteristic sets it apart from the finite field used in the traditional approach to RSSSs.  
A naive approach (see Fig. \ref{fig:scheme}(a)) is to map each element in a finite field to a probability vector, and then to employ RSSS over the field. To recover the secret, each share is mapped from the probability vector back to the finite field element. Since the composite DNA alphabet size grows cubically with the resolution $q$, the finite field needs to be large enough to achieve a high coding rate, which imposes demands on computational resources for the mapping and the encoding/decoding processes.
More importantly, within the traditional methodology, decoding of the secret necessitates sequencing and subsequent reconstruction of each available share \footnote{While it is possible to sequence multiple shares simultaneously using next-generation sequencing, e.g., Illumina sequencing, it requires adding markers on the DNA strands to distinguish the shares. As a result, the naive approach still imposes high operational complexity for information recovery.}.
This inevitably contributes to an escalation in recovery time, cost, and complexity.

The main contribution of the paper is a ramp secret sharing scheme for composite DNA storage systems (see Fig \ref{fig:scheme}(b)). Unlike the naive finite-field-based approach, it directly encodes and decodes over probability vectors.
Notably, this innovative scheme streamlines the process by eliminating the need for individually sequencing and reconstructing each share. 
Instead, it employs a mixture procedure, allowing for the direct recovery of the secret information, and reduces the sequencing requirement by a factor of $k/L$ (for example, in regular secret sharing $L=1$ and this factor equals $k$), thereby enhancing the efficiency in composite DNA storage applications.

In this paper, we first introduce a novel mixture operation for composite DNA letters that is both cost-effective and physically straightforward. Our mixture operation enables addition and multiplication (with positive or negative coefficients) of probability vectors. Leveraging this operation, we construct the Asymptotic Ramp Secret Sharing Scheme (ARSSS) for scalar probabilities, outlining the necessary rank conditions of the generator matrix for both information recovery and security. We start with an example to illustrate the construction and demonstrate the asymptotic information-theoretic data security achievable with a large alphabet size. We then provide a detailed proof of the generality of the recovery and the security properties. Next, we extend our construction from scalar probabilities to probability vectors and modify the recovery and secrecy proof to accommodate additional alphabet constraints. In the end, we examine, for any fixed resolution $q$, the existence of the generator matrix and define the objectives of  minimizing operational complexity and information leakage. To address these goals, we present practical and efficient constructions. We further adapt our approach to a scenario where the stored secret exceeds the data capacity of a single vessel. In this case, we utilize array codes to produce multiple vessels per share, and demonstrate the effectiveness in terms of the operational complexity and the information leakage.

{\bf Organization.} We present the problem and the definitions in Section \ref{sec: prob}. Then we
present the novel mixture operation for probability vectors in Section \ref{sec: operations}.
In Section \ref{sec:scheme}, we provide the ARSSS for scalar probabilities and probability vectors. Moreover, the existence of generator matrices, the practical matrix constructions, and analyses are shown in Section \ref{sec: const}, followed by conclusions in Section \ref{sec:conclusion}.

{\bf Notation.} A \emph{probability vector} is viewed as a single symbol with $m$ probability values, which together sum to the resolution. A \emph{probability sequence} is a length-$n$ tuple where each element is a probability vector. Real vectors, real matrices, and probability sequences are denoted by capital letters. Probability vectors and real numbers are denoted by lower-case letters. 
For a real matrix $G$, $|G|$ denotes the matrix of the same size whose elements are the absolute values of the elements of $G$ (Note that $|G|$ does not represent the determinant of $G$). Transposition of a matrix $G$ is denoted by $G^T$. We also write $G=(g_{i,j})$ to denote that the $(i,j)$-th element of the matrix is $g_{i,j}$. 
$\binom{n}{k}$ denotes the binomial coefficient $n$ choose $k$. For a positive integer $i$, denote $[i]=\{1,2,\dots,i\}$. For two integers $i<j$, denote $[i,j]=\{i,i+1,\dots,j\}$. For random sequences $S$ and $Y$, $H(S)$ denotes the entropy, and $H(S | Y)$ denotes the conditional entropy of $S$ given $Y$.

\section{Problem statement}\label{sec: prob}
In this section, we establish the definitions of probability sequences and ramp secret sharing.
\begin{definition}[Probability sequence] The \emph{probability sequence} $S=(s_1,\dots,s_n)^T$ has $n$ symbols or \emph{probability vectors}, and each symbol  $s_i=(s_{i1},\dots,s_{im}), i \in [n],$ has $m$ (scaled) \emph{probability values} satisfying
\begin{align}
&0 \le {s}_{i,j} \leq q, j\in[m], \label{eq:prob1}\\
&\sum_{j=1}^{m}{s}_{i,j}=q, \label{eq:prob2}
\end{align}
where $q$ is called the \emph{resolution}. Unless stated otherwise, $q$ is assumed to be a positive integer representing the complexity of reading/writing symbols. Although $q$ is also allowed to be a positive real number. The summation of probabilities equals 1, thus, $\sum_{j=1}^{j=m}\frac{s_{i,j}}{q}=1$. 
The set of all symbols  satisfying \eqref{eq:prob1} and \eqref{eq:prob2} is denoted by $\mathcal{Q}_{q, m}$. One can easily check that its size is $|\mathcal{Q}_{q, m}|=\binom{q+m-1}{m-1}$. The set of sequences of length $n$ is denoted as $\mathcal{Q}_{q, m}^n$. 
A \emph{uniform probability} sequence $U=(u_1,\dots,u_n)$ has $n$ symbols, and each symbol is the uniform probability vector with $m$ elements 
\begin{align}
    u_i=(\frac{q}{m},\dots,\frac{q}{m}). \label{eq:uniform}
\end{align}

\end{definition}

To simplify the notations, we often assume that $q$ is a multiple of $m$. 
In a broader context, the above definitions can all be adapted to varying resolutions, namely, we can assume the $i$-th symbol has resolution $q_i$. 

For example, when $m=4$, each probability vector $s_i=(s_{i1},s_{i2},s_{i3},s_{i4})$ is a composite DNA letter, corresponding to the nucleotides $\{A,C,G,T\}$. When $m=2$, we represent $s_i$ by the \emph{scalar probability} $s_{i1}$, and omit the second coordinate $s_{i2}=q-s_{i1}$. A scalar probability can be implemented by using only two types of nucleotides, e.g., $\{A,G\}$.

Upon establishing the definitions for probability sequences,  we can define $(k, L, n)$ asymptotic ramp secret sharing scheme (ARSSS) for probability sequences. 
\begin{definition}[$(k, L, n)$ ARSSS]\label{def:RSSS}
Assume a probability sequence $S=(s_1,s_2,\dots,s_L) \in \mathcal{Q}_{q, m}^L$ is the secret and $n \ge k \ge L$ are positive integers.
The scheme generates $n$ shares, denoted by a length-$n$ probability sequence $Y=(y_1,y_2,\dots,y_n)$. The resolutions of the shares may not be the same as $q$.
We define two conditions for the shares.

{\bf Recovery condition.} Denote any $k$ shares from $Y$ by $Y'$. Then $Y'$ gives all the information about $S$:  
\begin{align} \label{def: rec}
    H(S|Y') = 0.
\end{align}

{\bf Secrecy condition.} Denote any $k-L+j$ shares from $Y$ by $Y''$. Then as $q \to \infty$, $Y''$ gives partial information about $S$ with a ratio of $\frac{L-j}{L}$ for $ j \in [0,L-1]$:
\begin{align}\label{def: sec}
    \lim_{q \to \infty} \frac{H(S|Y'')}{H(S)} = \frac{L-j}{L}.
\end{align}
We say that a scheme is an asymptotic ramp secret sharing scheme (ARSSS) if the shares satisfy \eqref{def: rec} and \eqref{def: sec}.
\end{definition}

The converse proof of the secrecy condition, $H(S|Y'') \le \frac{L-j}{L}H(S)$, is shown in Theorem 1 of \cite{yamamoto}, where both $S$ and $Y$ belong to the same alphabet. We obtain a converse for arbitrary alphabet of $Y$ in Theorem \ref{thm:converse}.

The secrecy condition can be achieved if we map probability vectors to finite field elements, and employ a ramp secret sharing scheme over a finite field, such as the linear scheme in \cite{yamamoto}. Here, for a large enough $q$, we can choose a finite field size no more than the alphabet size $|\mathcal{Q}_{q,m}|$ and the reduction of the code rate can be negligible. However, the decoding requires $k$ sequencing operations to read the shares individually, which will be improved to $L$ by our proposed scheme in Section \ref{sec:scheme}.

We remark that our scheme is not linear, but facilitates simple encoding and recovery.

\section{Novel operations for probability vectors}\label{sec: operations}
In this section, we introduce a mixture operation, a cost-effective and physically straightforward procedure applicable to probability vectors. In addition, we introduce a novel mathematical operation for probability vectors, termed circle multiplication. These operations form the basis for our optimal ARSSS constructions. 

Let $g$ be a positive real number and $x=(x_1,\dots,x_m)$ be a probability vector. Then we write $gx = (gx_1,\dots,gx_m)$ as the product probability vector with resolution $gq$. 
For two probability vectors $x=(x_1,\dots,x_m), y=(y_1,\dots,y_m)$ with resolutions $q_1, q_2$, we write $x+y = (x_1+y_1,\dots,x_m+y_m)$ as the sum probability vector with resolution $q_1+q_2$. Similarly, for $G$ an $n \times k$ matrix with real positive entries and $X$ a probability sequence of length $k$, we write $GX$ as the probability sequence where $G$ is multiplied to every of the $m$ probability coordinates of $X$.

The addition and multiplication (with positive coefficients) of probability vectors can be implemented by the DNA mixture procedure, which is similar to the procedure used in \cite{mix0,mix1,mix2} for DNA computation. Briefly speaking, we get the samples from some given vessels, take the desired portions, and mix the samples to obtain new probabilities. The procedure is formally described below. 

 {\bf Mixture operation.}\label{de: mix}
    Formally, consider $k$ vessels where the $i$-th vessel represents probability vector $x_i=(x_{i1},\dots,x_{im})$, and has resolution $q_i$, for $i\in [k]$. Denote $X=(x_1,\dots,x_k)^T$ the probability sequence for these $k$ vessels. Let $G=(a_1,\dots,a_k)$ denote a positive real row vector of length $k$. Assume the desired probability after the mixture is
    \begin{align}
    y = GX = \sum_{i=1}^k g_i x_i, \label{eq: physical}
    \end{align}
    with resolution $q_y = \sum_{i=1}^k g_iq_i$. 
    Take $g_iq_i$ units of samples from the $i$-th vessel, for $i \in [k]$, and then mix them to obtain a mixture sample.
    Next we show that the desired probability vector $y$ is achieved.
    Denote the $m$ probability values of $y$ as $y=(y_1,\dots,y_m)$ and we wish to obtain a DNA sample such that for all $j \in [m]$, the probability of the $j$-th type of nucleotide is 
    \begin{align}
        \frac{y_j}{q_y} = \frac{1}{q_y} \sum_{i=1}^k g_i x_{i,j}, j \in [m].
    \end{align} 
    Note that the total amount of samples is $\sum_{i=1}^k g_iq_i$ units, and in the $i$-th vessel, the probability of the $j$-th type of nucleotide is $\frac{x_{i,j}}{q_i}$ for $i\in [k], j \in [m]$. Hence, in the mixture sample, the probability of the $j$-th type of nucleotide is 
    \begin{align}
    \frac{1}{\sum_{i=1}^k g_iq_j}\sum_{i=1}^k g_i q_i \frac{x_{i,j}}{q_i} = \frac{y_j}{q_y},
    \end{align}
    as desired.

    In practice, each vessel contains $M$ probability vectors arranged in multiple composite DNA sequences in order to represent a given data set, for some large integer $M$, and the above mixture achieves \eqref{eq: physical}  simultaneously for all the $M$ symbols.

Our ARSSS construction is similar to the linear methods over finite field \cite{yamamoto}, where we use a generator matrix and the corresponding decoding matrix to create shares and recover the secret. In the generator matrix and/or the decoding matrix, the presence of negative elements is inevitable, but negative probability vectors do not exist. To facilitate a secret sharing process allowing negative elements, the concept of a "negative" probability vector is introduced.

\begin{definition}(Restricted probability vector and "negative" probability vector):
A \emph{restricted probability vector} is defined to be a probability vector $s=(s_{1},\dots,s_{m})$ with resolution $q$, whose elements are in the range $[0,\frac{2q}{m}]$. Let $u=(\frac{q}{m},\dots,\frac{q}{m})$ be the corresponding uniform probability vector.
The \emph{negative probability vector} is defined as $s_{-}= 2u - s = (\frac{2q}{m}-s_{1},\dots,\frac{2q}{m}-s_{m}).$ 
Hence $s_{-}$ is also a restricted probability vector. 
The set of positive (or negative) restricted probability vectors is denoted by $\mathcal{Q}_{q, m}'$, whose size can be shown to be 
\begin{align}
    |\mathcal{Q}_{q, m}'|= \binom{q+m-1}{m-1}-\sum_{i=1}^{j} \binom{m}{i} \binom{q- (\frac{2q}{m}+1)i +m-i}{m-i+1},
\end{align}
where $j= \lfloor \frac{qm}{2q+m} \rfloor $. 
 
\end{definition}

For example, when $m=2$, a scalar probability $x=x_1 \in [0,q]$ has the negative scalar probability $x_{-}= q-x_1$. 
The alphabet $\mathcal{Q}_{q, 2}'$ is identical to $\mathcal{Q}_{q, 2}$. 
When $m=4,$ a DNA composite letter $x=(x_1,x_2,x_3,x_4),x_i\in [0,\frac{q}{2}], i\in [4],$ has the negative DNA composite letter $x_{-}= (\frac{q}{2}-x_1,\frac{q}{2}-x_2,\frac{q}{2}-x_3,\frac{q}{2}-x_4)$. The size of $\mathcal{Q}_{q, 4}'$ is calculated as
\begin{align}\label{eq: Q'}
    |\mathcal{Q}_{q, 4}'|= \binom{q+3}{3}-\binom{4}{1} \binom{\frac{q}{2}+2}{3}.
\end{align}

In our secret sharing scheme, unless otherwise stated, all the probability vectors are restricted probability vectors. 

\begin{remark}
    The generation of the negative probability vector is similar to the method of complements. The elements of the negative probability vector are $\left(\frac{2q}{m}\right)$'s complement of the elements of the positive probability vector. 
\end{remark}

Now we are ready to define the multiplication of an arbitrary real number and a probability vector.

\begin{definition}[Circle multiplication]\label{def: op}
Let $g$ be a real number, and $x$ be a probability vector of length $m$ with resolution $q$. Each probability value of $x$ is restricted in $[0, \frac{2q}{m}]$. The associated uniform vector is $u=(\frac{q}{m}, \dots, \frac{q}{m})$. Define the \emph{circle multiplication} as
\begin{align}
y &= g \otimes x \\
&= \begin{cases}
      g x,& \text{if } g\geq 0 \\
       |g| x_{-} = -g 2u + g x, & \text{if } g <0
   \end{cases} \label{eq:abs_sum}\\
   &=gx - gu + |g| u. \label{eq:circle1}
\end{align}
The resolution of $y$ is $|g|q$.

Consider a real $n \times k$ matrix $G=(g_{i,j})$ and a length-$k$ probability sequence $X=(x_1,\dots,x_k)^T$ with resolution $Q=(q_1,\dots,q_k)^T$. The corresponding uniform probability sequence is denoted by $U = (u_1,\dots,u_k)^T$, where $u_i=(\frac{q_i}{m},\dots,\frac{q_i}{m})$, for $i \in [k]$. 
Similar to \eqref{eq:circle1}, the circle multiplication of $G$ and $X$ is defined to be
\begin{align}\label{eq: mv_mul}
    Y=G \otimes X =GX-GU+|G|U,
\end{align}
where $|G|$ means that all the elements in $G$ are changed to their absolute value.
One can verify that the $i$-th entry of $Y$ is
\begin{align}
    y_i = \sum_{j=1}^{k} g_{i,j}  \otimes x_j.
\end{align}
From \eqref{eq:abs_sum}, we get that $Y$ has resolution 
\begin{align}\label{eq: res}
    |G|Q,
\end{align}
and the corresponding uniform probability sequence is
\begin{align}
    U' = |G|U. \label{eq:uniform_circle_mul}
\end{align}

\end{definition}

\begin{remark}
When the coefficient $g=0$, the circle multiplication can be calculated using either formula in \eqref{eq:abs_sum}. For simplicity, we treat $0$ as a ``positive'' coefficient.
Moreover, given a probability vector $x$, its resolution $q$ can be computed by summing up its $m$ values, and its associated uniform probability vector $u=(\frac{q}{m},\dots,\frac{q}{m})$ can be computed, too.
Therefore, in the definition of circle multiplication in \eqref{eq:circle1}, we treat $u$ as a known constant vector. Similarly, in \eqref{eq: mv_mul}, $U$ is treated as known. 
\end{remark}

\begin{example}[Vector-vector circle multiplication] \label{ex: vv_mul}
There are three scalar probabilities $X=(x_1,x_2,x_3)^T=(2,4,2)^T$, with resolution $Q=(q_1,q_2,q_3)=(6,8,10)^T$. 
Recall that for scalar probabilities or $m=2$, the second probability elements are omitted. So in this example, $X$ is a shorthand representation of the probability sequence $((2,4),(4,4),(2,8))^T$.
The uniform probabilities are $U=(u_1,u_2,u_3)=(3,4,5)^T$. Consider a real vector $G=(1,-2,2)$. Then 
\begin{align}
   &y = G \otimes X =GX -GU + |G|U\\
   & = (1,-2,2)
   \left(\begin{matrix}
       2\\
       4\\
       2\\
   \end{matrix}\right)
    -(1,-2,2)
   \left(\begin{matrix}
       3\\
       4\\
       5\\
   \end{matrix}\right)
   +(1,2,2)
   \left(\begin{matrix}
       3\\
       4\\
       5\\
   \end{matrix}\right)\\
   & = 14.
\end{align}
It can be also computed from \eqref{eq:abs_sum} as
\begin{align}
    y =& (1, -2, 2) \otimes 
     \left(\begin{matrix}
       2\\
       4\\
       2\\
   \end{matrix}\right)\\
    =& (1,2,2)  
    \left(\begin{matrix}
       2\\
       2u_2 - 4\\
       2\\
   \end{matrix}\right)
= 14. \label{eq:example18}
\end{align}

From \eqref{eq:example18}, the circle product $y$ can be achieved by the mixture procedure, but with the additional requirement that the negative probabilities are also stored as DNA samples.
We simply mix $q_1$ units of samples corresponding to $x_1$, $2q_2$ units of samples corresponding to the negative probability $x_{2-}$, and $2q_3$ units of samples corresponding to $x_3$. 

\end{example}

Next we show that the negative probability vector $y_{-}$ can be obtained from the circle multiplication of $g$ and $x_{-}$.
\begin{lemma}
    Let $x$ be a probability vector and $y=g\otimes x$.  Then
\begin{align}
  y_{-} = g \otimes x_{-} .\label{eq:neg_generation}
\end{align}
\end{lemma} 
\begin{IEEEproof}
    Let $x$ be a probability vector corresponding to the uniform probability vector $u$. Then $y=g \otimes x$ is associated with the uniform probability vector $|g|u$, and its negative probability vector is 
\begin{align}
y_{-} &=2|g|u - y \\
&= -gx+gu+|g|u \\
&= -g \otimes x \\
&= g(2u-x)-gu+|g|u \\
&= g \otimes x_{-}  .  \label{eq:negative_y}
\end{align}
where we used \eqref{eq:circle1} and $x_{-}=2u-x$.
\end{IEEEproof}

In general, for probability sequence $Y=G \otimes X$, we can similarly prove that
\begin{align}
  Y_{-} = G \otimes X_{-} .
\end{align}

\begin{remark}[Encoding method] \label{rmk:encoding_method}
Having introduced the circle product, we are now prepared to outline the method to generate $Y=G \otimes X$ and $Y_{-}$, which will be shown to be the encoding step of secret sharing. 
First, we synthesize both positive $X$ and negative $X_{-}$. 
If using certain synthesis method, such as degenerate bases\cite{choi2019}, terminal deoxynucleotidyl transferase\cite{lu2022enzymatic}, composite motifs \cite{composite_motifs}, and short DNA synthesis \cite{shortmer}, where different portions of nucleotides are added for each position, we can simultaneously generate both positive and negative probabilities. For example, assume $m=4$, and we wish to generate the probability vectors $x=(x_A,x_C,x_G,x_T)$ and $x_{-}$. We can take $\frac{q}{2}$ portions of $A$, put $x_A$ portions to the $x$ vessel, and the remaining $\frac{q}{2}-x_A$ portions to the $x_{-}$ vessel. The same process is repeated for the other nucleotide types.
Once we have both positive $X$ and negative $X_{-}$, we utilize the mixture procedure to create $Y=G \otimes X$ and $Y_{-}=G \otimes X_{-}$. 
If all coefficients in the matrix $G$ are positive, we only need positive $X$ to generate $Y=G \otimes X$, and the number of synthesis operations is $k$. To generate both $Y=G \times X$ and $Y_{-}=G \otimes X_{-}$, we need $2k$ synthesis operations. 
\end{remark}

In the following, Lemma \ref{lem: decode} states that the circle multiplication can be inverted if $G$ is an invertible matrix, similar to regular matrix-vector multiplication.
Then we will use Example \ref{ex: decode} to illustrate it.
\begin{lemma}\label{lem: decode}
    Let $G$ be an invertible matrix of size $k \times k$, $X$ be a probability sequence of length $k$, and $Y = G \otimes X$. Denote by $U$ the uniform probability sequence corresponding to $X$. Then
    \begin{align}\label{eq: decode}
         X = G^{-1} \otimes Y + U -  |G^{-1}||G|U .
    \end{align}
    Treating $U,G^{-1},G$ as known constants, $X$ can be recovered given $Y$, namely, 
    \begin{align}
        H(X|Y)=0. \label{eq:recovery_X_Y}
    \end{align}
\end{lemma}
\begin{IEEEproof}
    Assume $G^{-1}$ is the inverse of $G$. According to \eqref{eq:uniform_circle_mul}, the uniform probability sequence associated with $Y$ is $U' = |G|U$. Then the right-hand side (RHS) of \eqref{eq: decode} can be expanded by following \eqref{eq: mv_mul}:
\begin{align}
    &RHS \nonumber\\
    =& G^{-1}Y - G^{-1}U' + |G^{-1}|U' + U -  |G^{-1}||G|U \\
    =& G^{-1} (GX-GU+|G|U)  - G^{-1}|G|U \nonumber\\ 
    & + |G^{-1}||G|U + U -  |G^{-1}||G|U \\
   =& X ,
\end{align} 
Equation \eqref{eq:recovery_X_Y} is an immediate result of Equation \ref{eq: decode}.
\end{IEEEproof}

\begin{example}\label{ex: decode}
Consider the scalar probability sequence $X =(2, 4)$ with resolution $Q=(8,8)^T$ and $U=(4,4)^T$. Let $G$ be the matrix
\begin{align}
\left[\begin{matrix}
       1&   1\\
         1&  -1
   \end{matrix}\right] .
\end{align}
Then the result of the matrix-vector circle multiplication is
\begin{align}
Y & = G \otimes X = GX+(|G|-G)U\\
&=\left[\begin{matrix}
       1&   1\\
         1&  -1
   \end{matrix}\right] \left(\begin{matrix}
       2\\
       4\\
   \end{matrix}\right) +  \left(\left[\begin{matrix}
       1&   1\\
         1&  1
   \end{matrix}\right] -\left[\begin{matrix}
       1&   1\\
         1&  -1
   \end{matrix}\right] \right)\left(\begin{matrix}
       4\\
       4\\
   \end{matrix}\right)\\
   &=\left(\begin{matrix}
       6\\
       6\\
   \end{matrix}\right).
\end{align}
The resolution of $Y$ is $|G|Q = (16,16)^T$ and the corresponding uniform probability is $U'=(8,8)^T$. 

Then we use $Y$ to retrieve the $X$ based on Equation \eqref{eq: decode}:
\begin{align}
    X &= G^{-1} \otimes Y + U -  |G^{-1}||G|U \label{eq:38}\\
   &= G^{-1} (Y - U') + |G^{-1}|U' + (I - |G^{-1}||G|)U \label{eq:39}\\
    &=  \left[\begin{matrix}
       \frac{1}{2}&   \frac{1}{2}\\
          \frac{1}{2}&  - \frac{1}{2}
   \end{matrix}\right]  \left(\begin{matrix}
       -2\\
       -2\\
   \end{matrix}\right)+ \left[\begin{matrix}
       \frac{1}{2}&   \frac{1}{2}\\
          \frac{1}{2}&  \frac{1}{2}
          \end{matrix}\right] \left(\begin{matrix}
       8\\
       8\\
   \end{matrix}\right)-\left(\begin{matrix}
       4\\
       4\\
   \end{matrix}\right)\\
    &=\left(\begin{matrix}
       2\\
       4\\
   \end{matrix}\right).
\end{align}
We note here that despite the fact that the resolution of $G^{-1} \otimes Y$ is $|G^{-1}||G|Q=(16,16)^T$, which is different from the resolution $Q$ of $X$, we can still successfully retrieve $X$ by \eqref{eq:38}.
\end{example}

\begin{remark}[Sequencing cost reduction]\label{rmk: retrieval}
In the following, we focus on the recovery of the first entry of $X$ given $Y=G\otimes X$, and assume $X$ and $Y=(y_1,y_2,\dots,y_k)^T$ are of length $k$.
For comparison, a naive strategy maps probabilities to finite field elements, performs normal matrix-vector multiplication $X=G^{-1}Y$ over the field, and necessitates the sequencing and reconstruction of every coordinate of $Y$. Thus, the naive strategy requires $k$ sequencing operations.

There are two methods to achieve the sequencing operation reduction based on the proposed circle multiplication. 
Denote by $y_{i-}$ the negative probability vector of $y_i$, and let $u_i$ be the uniform probability associated with $y_i$.
Assume the first row of $G^{-1}$ is $A=(a_1,a_2,\dots,a_k)$. Note that there are two ways to obtain the circle multiplication:
\begin{align}
    & A \otimes Y \nonumber \\
    =& \sum_{i: a_i \ge 0} a_i y_i + \sum_{i: a_i < 0} |a_i| y_{i-} \label{eq:mix_1}\\
    =& \sum_{i: a_i \ge 0} a_i y_i - \sum_{i: a_i < 0} |a_i|   y_{i} + \sum_{i: a_i < 0} 2 |a_i| u_i  \label{eq:mix_2}.
\end{align}

Method (i). We synthesize $2k$ times to generate all probability vectors in $Y$ and $Y_{-}$ according to Remark \ref{rmk:encoding_method}. In decoding of the first entry of $X$, we simply mix the shares from $Y$ and $Y_{-}$ according to \eqref{eq:mix_1}, and then sequence once to retrieve $A \otimes Y$. Finally, the first entry of $X$ is obtained by \eqref{eq:38}. In this procedure, we can achieve a $k$-fold reduction on the sequencing operations but require twice the synthesis cost. 

Method (ii). We only synthesize $k$ times to generate $Y$ according to Remark \ref{rmk:encoding_method}, assuming all coefficients in $G$ are positive. During decoding, we obtain two different mixture samples. One sample is called the positive sample, which corresponds to the positive coefficients, i.e., $\sum_{i: a_i \ge 0} a_i y_i$ in \eqref{eq:mix_2}. The other is called the negative sample which mixes the shares associated with the negative coefficients, i.e., $\sum_{i: a_i < 0} |a_i|   y_{i}$ in \eqref{eq:mix_2}. Then sequence these two samples and calculate the result of \eqref{eq:mix_2}. Finally, use \eqref{eq:38} to compute the first entry of $X$. In this procedure, we can achieve a $\frac{k}{2}$-fold reduction on the sequencing operations without extra synthesis cost.
\end{remark}

\section{Asymptotic ramp secret sharing schemes for probability vectors}\label{sec:scheme}

We are now prepared to introduce an ARSSS for scalar probabilities. 
This will serve as a foundation before we explore the scheme designed for composite DNA letters.
Similar to Remark \ref{rmk: retrieval}, our scheme achieves a reduction in sequencing operations compared to the finite-field-based schemes. 

\begin{construction}[$(k,L,n)$ ARSSS]\label{cnstr:RSSS}
    Let $n \ge k \ge L$.
    Fix a secret $S=(s_1,\dots, s_L)^T$, where each element $s_i, i \in [L]$ is an integer uniformly selected from $[0,q]$. We generate the auxiliary sequence $X = (x_1,\dots,x_k)^T$, such that  $x_i=s_i, i\in [0, L]$ and $x_j, j\in [L+1, k]$ is uniformly selected from $[0,q]$.  
    Let $G=(g_{i,j})$ be  an integer matrix with dimension $n \times k$. It satisfies the following {\bf Rank Conditions}: (i) any $k$ rows of $G$ are full rank, and (ii) any submatrix of $G$ consisting of any $k-L$ rows and the last $k-L$ columns should be full rank. Encode $X$ into $n$ shares $Y = G \otimes X$.
\end{construction}

The existence and the construction of the generator matrix $G$ are detailed in Section \ref{sec: const}.

In terms of secrecy, we need to find how many solutions or how much information about the secret can be retrieved from a given set of shares.
It is evident that the number of solutions, if there are any, to $Y=GX$ equals $q^{i}$ when all elements of $X$, $G$, and $Y$ are in a finite field of size $q$, where $i$ equals the number of unknowns minus the rank of $G$. However, the number of solutions is not easy to describe if the elements are not in a finite field. The challenge arises from the possibility that certain solutions are not valid probabilities. 

A useful tool for our calculations is the following lemma from reference \cite{Zippel1993} concerning solutions to linear Diophantine equations by computing the Smith normal form of the equations, which are linear equations with integer coefficients and unknowns \cite{1969diophantine,2006diophantine,2001solving,2010diophantine}. In this lemma, a unimodular matrix is defined to be a square integer-valued matrix with determinant $1$ or $-1$. Moreover, $\mathbb{Z}$ denotes the set of integers. 

\begin{lemma}\label{lem: Diophantine}
    Consider the linear Diophantine equations
\begin{align}
    Y=GX,
\end{align}
where $G \in \mathbb{Z}^{l \times k} 
 (l<k)$ is full rank and  $X \in \mathbb{Z}^{k \times 1}$. 
There exist unimodular matrices $V_1\in \mathbb{Z}^{k \times k}$ and $V_2 \in \mathbb{Z}^{l \times l}$ such that
\begin{align}
    B=(b_{i,j})= V_2GV_1,
\end{align}
is a diagonal matrix with nonzero $b_{i,i}$, for all $i \in [l]$.
Let $D= (d_1,\dots,d_l)^T = V_2Y$, and $M = [\frac{d_1}{b_{1,1}},...,\frac{d_l}{b_{l,l}}, r_{l+1},..., r_k ]^T$, where $r_{l+1},..., r_k$ are arbitrary integers.
The equations have solutions if and only if $b_{i,i}$ divides $d_i$ for $i \in [l]$, which are
\begin{align}    
X = V_1M.
\end{align}
\end{lemma}

The following example of Construction \ref{cnstr:RSSS} applies Lemma \ref{lem: Diophantine} and computes the remaining secret uncertainty given a set of unauthorized shares.
\begin{example}[$(2,1,2)$ ARSSS]
\label{ex: 1}
First form the auxiliary vector $X=(s,t)^T$, where $s \in [0,q]$ is the secret and $t$ is a uniformly random integer in $[0,q]$. 
The shares are $Y=(y_1,y_2)^T= G \otimes X$ where
\begin{align}
  G =  \left(\begin{matrix}
      G_1\\
      G_2\\
  \end{matrix}\right) =
\begin{bmatrix}
  1 & 1\\
  1&  -1 
\end{bmatrix}.
\end{align}

The encoding and decoding are realized by the mixture operations. 
The recovery condition \eqref{def: rec} was shown in Example \ref{ex: decode}. In the following, we focus on showing the secrecy condition $\frac{H(s|y_1)}{H(S)} \to 1$. In words, as the alphabet size grows to infinity, the uncertainty of the secret $s$ given just one share asymptotically approaches the entropy of the secret.

By Definition \ref{def: op}, $y_1=G_1 \otimes X = G_1 X$ since $G_1$ has positive elements. 
Given $y_1$, $0 \le y_1 \le 2q$, denote $N(y_1)$ as the number of $s$ values satisfying the equation $y_1 = G_1 (s,t)^T$. 
By Lemma \ref{lem: Diophantine}, one can find $V_2=1$, $V_1=\begin{bmatrix}
    1 &-1\\
    0& 1
\end{bmatrix}$, and $(s,t) = (y_1-r,r)$, where $r$ is an arbitrary integer. 
Thus given $y_1$, every $s$ value uniquely maps to one $(s,t)$ pair, so $N(y_1)$ equals the number of solutions of $(s,t)$, and $\sum_{i=1}^{2q} N(i)$ equals the number of pairs $(s,t)$, which is $(q+1)^2$. Moreover, given $y_1$, every $s$ value is equally likely, since $s,t$ are uniform in $[0,q]$. Thus,
\begin{align}
    &P(y_1=i) = \frac{N(i)}{\sum_{j=1}^{2q} N(j)} = \frac{N(i)}{(q+1)^2}, \label{eq:44}\\
    &P(s=j|y_1=i) = \frac{1}{N(i)}, \text{for any $j$ s.t. $i=G(j,t)^T$}. \label{eq:45}
\end{align}
Since $(s,t)$ must be in $[0,q]$, 
\begin{align}
    &0\leq s=y_1-r\leq q,\\
    &0\leq t=r \leq q.
\end{align}
Therefore, 
\begin{align}
   N(y_1) &= \begin{cases}
        y_1+1, & \text{if } y_1 \in [0,q],\\
        2q-y_1+1, & \text{if } y_1 \in [q+1,2q].
    \end{cases}\label{eq:49}
\end{align}
Combining \eqref{eq:44} \eqref{eq:45} and \eqref{eq:49}, we get
\begin{align}
    &H(s|y_1) \\
    =& \sum_{i=0}^{2q} P(y_1 = i) H(s | y_1=i) \label{eq:50}\\
    =& \sum_{i=0}^{2q} \frac{N(i)}{(q+1)^2} \log_2 N(i) \label{eq:51}\\
    =&\frac{1}{(q+1)^2} \left( 2\sum_{i=1}^{q+1} i \log_2 i - (q+1) \log_2 (q+1)\right)\\
    =&\frac{1}{(q+1)^2} \left(2 \log_2 f(q+1) - (q+1)\log_2 (q+1)\right),
\end{align}
where $f(q+1)=\prod_{i=1}^{q+1} i^i$ is the hyperfactorial. 
Based on the above equation, we show the ratio of $ H(s|y_1)$ and $H(s)$ for different $q$ in Table \ref{tab: comparison}.
We can see that the ratio between $ H(s|y_1)$ and $ H(s)$ increases as $q$ goes up. For a large alphabet size $q$, by the asymptotic approximation \cite{alabdulmohsin2018summability} of $f(q+1) \approx A (q+1)^{\frac{1}{2}(q+1)^2+\frac{1}{2}(q+1)+\frac{1}{12}}e^{-\frac{(q+1)^2}{4}}$, where $A \approx 1.28$ is the Glaisher's constant, we get
\begin{align}
    \lim_{q\to \infty} \frac{H(s|y_1)}{H(s)} = \lim_{q \to \infty}\frac{\frac{2}{(q+1)^2} \cdot \frac{1}{2}(q+1)^2 \log_2 (q+1)}{\log_2 (q+1)} = 1.
\end{align}
Therefore, the asymptotic secrecy condition \eqref{def: sec} is proved when $Y''=y_1$. When $Y''=y_2$, the secrecy condition follows similarly. The only difference is that the pair $(s,t)$ is changed to the pair $(s,t_{-})$ based on circle multiplication. But the number of solutions remains the same, because $t_{-}$ is also uniformly distributed in $[0,q]$.

\begin{table}
    \centering
    \begin{tabular}{c|c|c|c|c}
    \hline
    $q $  &   4  &   8    & 12  & 16\\
    \hline
$    \frac{H(S|y_1)}{H(S)} $&    0.7084 &   0.7773  &   0.8072   & 0.8247\\
\hline
    \end{tabular}
    \caption{Ratio between $ H(s|y_1)$ and $ H(s)$}
    \label{tab: comparison}
\end{table}
\end{example}

We state in Theorem \ref{thm: RSSS} that the observation of asymptotic optimality from the above example is generalizable to any $(k, L, n)$ and any generator matrix $G$ of Construction \ref{cnstr:RSSS}. 

\begin{theorem}\label{thm: RSSS}
    Construction \ref{cnstr:RSSS} for scalar probabilities is an ARSSS satisfying the recovery and the secrecy conditions. 
\end{theorem}

Theorem \ref{thm: RSSS} directly follows from Lemmas \ref{lem: recovery} and \ref{lem: secrecy} stated below.
\begin{lemma}\label{lem: recovery}
    Given any $k$ shares $Y'$ of the code in Construction \ref{cnstr:RSSS}, the recovery condition \eqref{def: rec} is satisfied, i.e., the original secret $S$ can be recovered.
\end{lemma}
\begin{IEEEproof}
 It holds because of Rank condition (i) and Lemma \ref{lem: decode}.
 \end{IEEEproof}

\begin{lemma}\label{lem: secrecy}
    Any $k-L+j$ shares $Y''$ of the code in Construction \ref{cnstr:RSSS} satisfies the asymptotic secrecy condition \eqref{def: sec}, for $j \in [0,L-1]$.
\end{lemma}

\begin{IEEEproof}
First, let us determine the entropy of $S,X$ and $Y''$.
The secret $S$ and the auxiliary sequence $X$ are uniform, and hence $H(S)=L\log_2(q+1)$, $H(X)=k\log_2(q+1)$. 
Without loss of generality, assume $Y''$ are the first $k-L+j$ shares.
Let $g_i$, $i \in [k-L-j]$, be the sums of absolute values in the rows of the generator matrix associated with the shares $Y''$. By \eqref{eq: res} the alphabet size of $Y''$ is $ \left(\prod_{i=1}^{k-L+j} g_i \right) (q+1)^{k-L+j}$. Denote $g=\prod_{i=1}^{k-L+j} g_i$, and then $H(Y'') \le  (k-L+j)\log_2 (q+1) + \log_2 g$. Note that $g$ is a constant independent of $q$.
Thus, the conditional entropy of $S$ given $Y''$ can be bounded as follows.
\begin{align}
   H(S|Y'') 
&= H(S|Y'') - H(S|Y'', X) \label{eq:70}\\ 
   & = I(S;X|Y'')\\
   &=H(X|Y'') - H(X|S,Y'')\\
 & = H(X|Y'') \label{eq:73}\\
 &= H(X,Y'') - H(Y'') \\
 &= H(X)-H(Y'') \label{eq:74}\\
 & \geq (L-j) \log_2 (q+1) - \log_2 g \label{eq:exact_lower}\\ 
 &= \frac{L-j}{L} H(S) - \log_2 g.
\end{align}
Here, \eqref{eq:70} follows by the fact that given $X$, the secret $S$ is uniquely determined, \eqref{eq:73} follows by the fact that given $S,Y''$, the auxiliary sequence $X$ is uniquely determined due to Rank condition (ii) in Construction \ref{cnstr:RSSS}, and \eqref{eq:74} is due to the fact that $X$ uniquely determines $Y''$. The asymptotic secret condition \eqref{def: sec} is proved by taking $q \to \infty$.
\end{IEEEproof}

Next, we introduce ARSSS for non-scalar probability vectors. We consider the case for DNA composite letters with $m=4$, but this method can also be easily modified for probability vectors with any length $m$. 

When $m=4$, each symbol encompasses four probability elements. Construction \ref{cnstr:RSSS} remains the same, except that all scalar probabilities are changed to probability vectors of length $m=4$, where each probability value is restricted in $[0,\frac{q}{2}]$, for some even resolution $q$.

\begin{theorem}\label{thm: 2}
    Construction \ref{cnstr:RSSS} for  probability vectors with $m=4$ is an ARSSS satisfying the recovery and the secrecy conditions.  
\end{theorem}
\begin{IEEEproof}[Sketch of Proof] 
We apply the same $n \times k$ generator matrix $G$ of Construction \ref{cnstr:RSSS} to each of the $4$ probability positions. Treating each probability value as one digit, the number of equations and the unknowns are quadrupled.
Denote $G^*$ as the resulting $4n \times 4k$ generator matrix for composite DNA letters. It is not hard to see that $G^*$ can be formed by repeating $G$ diagonally four times and
\begin{align}\label{eq:newg}
    G^* = \begin{bmatrix}
        G &0 &0& 0\\
        0& G& 0 &0 \\
        0& 0& G &0\\
        0 &0 &0 &G
    \end{bmatrix},
\end{align}
where $G$ is an integer matrix with dimension $n \times k$ satisfying the rank conditions for scalar probabilities, defined in Construction \ref{cnstr:RSSS}. The $0$ in Equation \eqref{eq:newg} is an all-zero  matrix with dimension $n \times k$. The corresponding auxiliary sequence $X$ is written such that the first coordinates of all probability vectors in $X$ come first, followed by the second coordinates of all probability vectors, and so on. Namely, $X=(x_{11},\dots,x_{k1},$ $x_{12},\dots,x_{k2},$ $x_{13},\dots,x_{k3},$ $x_{14},\dots,x_{k4})^{T}$. Recall that $x_i=(x_{i1},x_{i2},x_{i3},x_{i4})$ is the $i$-th probability vector in $X$. 
The first $L$ probability vectors of $X$ are the secret, and the last $k-L$ probability vectors of $X$ are uniformly random probability vectors.
Similarly, the shares $Y=G^* \otimes X$ also group the coordinates of the probability vectors together.

The grouping of the coordinates does not affect the property of the circle multiplication. Then the recovery condition proof is the same as Lemma \ref{lem: recovery}, which requires Rank condition (i) and Lemma \ref{lem: decode}. The secrecy condition proof follows the same line of Lemma \ref{lem: secrecy}. The difference is in the alphabet size in Equation \eqref{eq:exact_lower}. 
In particular, $H(X)=\log_2 |\mathcal{Q}_{q, 4}'|$, $H(Y) \le \log_2 |\mathcal{Q}_{g_i q.4}'|$, where the sizes of $\mathcal{Q}_{q, 4}'$ and $\mathcal{Q}_{g_i q.4}'$ can be calculated using \eqref{eq: Q'}. 
When $q \to \infty$, the asymptotic secret condition \eqref{def: sec} still holds. Therefore, Construction \ref{cnstr:RSSS} is an ARSSS for composite DNA letters. 
\end{IEEEproof}

\begin{remark}
Even though the $m$ probability values are dependent, as they sum to $q$, the recovery and secrecy conditions are still met due to the aforementioned argument. In other words, representing information with probability vectors, as opposed to using finite field elements, effectively maintains the necessary conditions for secret sharing. 
\end{remark}

In the following, we use an example to illustrate the ARSSS construction for DNA composite letters, and show its recovery and secrecy conditions. Moreover, we calculate the secret information uncertainty given $k-L+j$ shares $Y''$ for a finite $q$.
Unlike the calculation of $H(S|Y'')$ for scalar probabilities, the number of solutions given $Y''$ should be calculated using $Y=G^* \otimes X$ in addition to some extra equations, which is that the sum of any probability vector in $X$ is $q$.

\begin{example}
Assume the setting is the same as Example \ref{ex: 1}, except that the symbols are probability vectors.

Note that $X=(s,t)^T$, $s=(s_1,s_2,s_3,s_4)^{T}$, $t=(t_1,t_2,t_3,t_4)^{T}$ and $y_1=(y_{11},y_{12},y_{13},y_{14})^{T}$. The uniform probability vector for $s,t$ is $(\frac{q}{2},\frac{q}{2},\frac{q}{2},\frac{q}{2})$. 
First, let us show the recovery condition \eqref{def: rec}, $H(s|y_1,y_2)=0$.
Equation $Y = G^* \otimes X$ can be written as 
\begin{align}
    &s_i + t_i = y_{1i}, i \in [4]\label{eq:s_plus_t}\\
    &s_i - t_i = y_{2i}-\frac{q}{2}, i \in [4] \label{eq:s_minus_t}
\end{align}
One can check that there are $8$ independent equations and $8$ unknowns. Therefore, the recovery condition \eqref{def: rec} is satisfied.
Moreover, summing all equations in \eqref{eq:s_plus_t} and \eqref{eq:s_minus_t}, we get
\begin{align}
    &s_1 + s_2 + s_3 + s_4 = q, \label{eq:sum1}
\end{align}
where we used the fact that the sum of probabilities in $y_1$ (and $y_2$ respectively) equals $2q$.
Similarly, the difference of \eqref{eq:s_plus_t} and \eqref{eq:s_minus_t} gives
\begin{align}
    &t_1 + t_2 + t_3 + t_4 = q.\label{eq:sum2}
\end{align}
We can see that the solutions to \eqref{eq:s_plus_t} \eqref{eq:s_minus_t} are indeed probability vectors.

Next, we compute the remaining secret given one share, $H(s|y_1)$.
The alphabet size of $y_1$ is $|\mathcal{Q}_{2q,4}'|=\binom{2q+3}{3}-\binom{4}{1}C_{q+2}^{3}$, and the alphabet size of $(s,t)$ is $|\mathcal{Q}_{q, 4}'|^2 = \left(\binom{q+3}{3}-\binom{4}{1}\binom{\frac{q}{2}+ 2}{3}\right)^2$. By an argument similar to \eqref{eq:exact_lower}, one can show that
\begin{align}
    H(s,t|y_1) 
 &  \ge \log_2 \frac{|\mathcal{Q}_{q, m}'|^2}{|\mathcal{Q}_{2q}'|}\\
 &=\log_2 |\mathcal{Q}_{q, m}'| + o(\log_2|\mathcal{Q}_{q, m}'|)\\
 &=H(s) +  o(\log_2|\mathcal{Q}_{q, m}'|).
\end{align}
Moreover, Rank condition (ii) and Equations \eqref{eq:sum1}, \eqref{eq:sum2} imply that $H(X|s,y_1)=0$. Similar to \eqref{eq:73} we can obtain 
$H(s|y_1)=H(s,t|y_1)$. Taking $q \to \infty$, the asymptotic secrecy condition is satisfied.

For finite $q$, the number of solutions of the secret $s$ given the single share $y_1$ is calculated as follows. 
Since each probability vector sums to $q$, we can ignore $s_4, t_4$ when writing the encoding equations. 
Equation \eqref{eq:s_plus_t} is rewritten as
\begin{align}
    \begin{pmatrix}
  1 & 1& 0 &0 &0 &0 \\
0 &1& 1& 0& 0& 0\\
0& 0& 1 &1 &0 &0\\
 0 &0 &0 & 0&1& 1  
    \end{pmatrix}    \begin{pmatrix}
    s_1\\
     t_1\\
s_2 \\
 t_2\\
s_3\\
 t_3
    \end{pmatrix} =  \begin{pmatrix}
    y_{11}\\
y_{12} \\
y_{13}\\
 y_{14}
    \end{pmatrix}.
\end{align}
Here, we know the rank of the coefficient matrix is $4$. Based on Lemma \ref{lem: Diophantine} and similar to Example \ref{ex: 1}, we have
\begin{align}\label{diophantine}
    \begin{bmatrix}
    s_1\\
     t_{1}\\
    s_2\\
      t_2\\
    s_3\\
        t_3
    \end{bmatrix} = V_1
    \begin{bmatrix}
    y_{11}-y_{12}+y_{13}\\
   y_{12}-y_{13}\\
   y_{13}\\
   y_{14}\\
   r_2\\
    r_3\\
    \end{bmatrix},
\end{align}
where $r_2,r_3$ are arbitrary integers and $V_1$ is
\begin{align}
    \begin{pmatrix}
 1	&0	&0	&0	&-1	&0\\
0	&1	&0	&0	&1	&0\\
0	&0	&1	&0	&-1	&0\\
0	&0	&0	&0	&1	&0\\
0	&0	&0	&1	&0	&-1\\
0	&0	&0	&0	&0	&1
       \end{pmatrix}.
\end{align}
Since $(s,t)$ must be in $[0,\frac{q}{2}]$, 
\begin{align}
    &0\leq s_1=y_{11}-y_{12} +y_{13} - r_2\leq \frac{q}{2},\\
    &0\leq t_1=y_{12}- y_{13} + r_2 \leq \frac{q}{2},\\
    &0\leq s_i = y_{1i}-r_i \leq\frac{q}{2}, i\in[2,3],\\
    &0\leq t_i=r_i \leq \frac{q}{2} , i\in[2,3].
\end{align}
Given $y_1=(y_{11},y_{12},y_{13},y_{14})^{T}$, every $s$ value uniquely maps to one $(s,t)$ pair. Therefore, the number of $y_1$ equals to the number of solution of $(s,t)$ pair. It is similar to Example \ref{ex: 1}, but $y$ only has one digit in Example \ref{ex: 1}. Combined with the summation restriction $\sum_{i=1}^{4 }y_{1i} = 2q$ and following similar steps as Equation \eqref{eq:51}, we can calculate $ H(s|y_1)$. We show the ratio of $ H(s|y_1)$ and $H(s)$ for different $q$ in Table \ref{tab: comparison2}. It can be seen that the ratio grows as $q$ increases.

\begin{table}[h]
    \centering
    \begin{tabular}{c|c|c|c|c}
    \hline
    $q $  &   4  &   8    & 12  & 16\\
    \hline
$    \frac{H(S|y_1)}{H(S)} $&    0.4757 &   0.5594  &   0.5972   & 0.6191\\
\hline
    \end{tabular}
    \caption{Ratio between $ H(s|y_1)$ and $ H(s)$}
    \label{tab: comparison2}
\end{table}
\end{example}

In Theorem \ref{thm:converse} we give the converse of the secret condition \eqref{def: sec}, which is different from Theorem 1 of \cite{yamamoto} in terms of the alphabet. In \cite{yamamoto}, it is assumed that the alphabet size of each entry of $S,X,Y$ is the same. Recall that the shares are denoted as $Y=(y_1,y_2,\dots,y_n)^T$. In the following theorem, we assume that all entries in $S$ are from the same alphabet $\mathcal{Q}_s$, all entries in $X$ belong to the same alphabet $\mathcal{Q}_x$, and each share $y_i$, $i \in [n]$ belongs to the alphabet $\mathcal{Q}_{y_i}$. Moreover, denote $Y_{1}^j=(y_1,y_2,\dots,y_j)^T$, for any $j \in [n]$. Note that our theorem holds for schemes over any alphabet, and is not restricted to probability vectors.

\begin{theorem}\label{thm:converse}
Consider a sequence of constructions with growing alphabet sizes satisfying the recovery condition \eqref{def: rec}, where the alphabet sizes satisfy $\frac{\log_2\left(\left|\mathcal{Q}_{y_i}\right| \right)}{\log_2\left(\left|\mathcal{Q}_s\right| \right)} \to 1$, as $\left|\mathcal{Q}_s\right| \to \infty$, for all $i\in [n]$.
For any $k-L+j$ shares $Y''$ of $Y$, $j \in [0,L]$, the following upper bound holds:
\begin{align}\label{eq:sec}
    \lim_{\left|\mathcal{Q}_s\right| \to \infty} \frac{H(S|Y'')}{H(S)} \le \frac{L-j}{L}.
\end{align}
\end{theorem}
\begin{IEEEproof}
Without loss of generality, assume $Y''=Y_1^{k-L+j}$. 
\begin{align}
&H(S|Y_1^k) = H(S, Y_1^k) - H(Y_1^k)\\
& \geq H(S, Y_1^{k-1} )- H(Y_1^{k-1})- H(y_k|Y_1^{k-1}) \\
&\geq H(S|Y_1^{k-1}) - H(y_k)\\
&\geq H(S|Y_1^{k-1}) - \log_2 \left| \mathcal{Q}_{yk} \right|.
\end{align} 
Continuing the same argument for $L-j$ steps, we get
\begin{align}
H(S|Y'') &= H(S|Y_1^{k-L+j})\\
&\leq H(S|Y_1^{k}) + \sum_{i=k-L+j+1}^{k} \log_2 \left|\mathcal{Q}_{y_i} \right|\\
&= \sum_{i=k-L+j+1}^{k} \log_2 \left| \mathcal{Q}_{y_i} \right|, \label{eq:sum_Q_y}
\end{align}
where the last step follows from the recovery condition \eqref{def: rec}.
Therefore, 
\begin{align}
    \frac{H(S|Y'')}{H(S)} &\le \frac{\sum_{i=k-L+j+1}^{k} \log_2 \left| \mathcal{Q}_{y_i} \right|}{L \log_2 \left| \mathcal{Q}_s \right| }   \label{eq:68} \\
    &\to \frac{L-j}{L},
\end{align}
as $\left|\mathcal{Q}_s\right| \to \infty$.
\end{IEEEproof}

Let us apply \eqref{eq:sum_Q_y} of Theorem \ref{thm:converse} to find the upper bound of $H(S|Y'')$, for an arbitrary construction with the same secret alphabet and share resolution as Construction \ref{cnstr:RSSS}. 
The alphabets for the secret and the share are $\mathcal{Q}_s  =  \mathcal{Q}_{q,m}'$, and $\mathcal{Q}_{y_i}=\mathcal{Q}_{g_i q,m}$, $i \in [n]$. Recall that $g_i$ is the sum of the absolute values in $i$-th row of the generator matrix $G$, and is a constant independent of $q$. Here, we set the secret alphabet to be exactly the same as Construction \ref{cnstr:RSSS} so that the amount of encoded secret is identical and the the comparison is fair. However, the shares are allowed to be all probability vectors instead of restricted probabilities, so as to allow potentially better coding schemes.

When $m=2$, $|\mathcal{Q}_{q,m}|=|\mathcal{Q}'_{q,m}|=q+1$. Therefore, $\left|\mathcal{Q}_s\right| = |\mathcal{Q}'_{q,m}|= q+1 $ and $\left|\mathcal{Q}_{y_i} \right| = |\mathcal{Q}_{g_i q,m}|=  g_i (q+1)$.  The remaining secrecy given the shares $Y''$ in Equation \eqref{eq:sum_Q_y} can be shown to satisfy
\begin{align}
    H(S|Y'') \le (L-j)\log_2(q+1)  + \sum_{i=k-L+j+1}^{k}\log_2 g_i. \label{eq:exact_upper}
\end{align}
Taking $q \to \infty$, and dividing both sides by $H(S)=L\log_2(q+1)$, we have Equation \eqref{eq:sec}.
Combining Lemma \ref{lem: secrecy}, we conclude that Construction \ref{cnstr:RSSS} has asymptotic optimal secrecy. 

When $m=4$, there is a difference between $\mathcal{Q}_{q,m}$ and $\mathcal{Q}'_{q,m}$. However, we will see that this difference does not affect the asymptotic optimality.
The secret alphabet size is $\left|\mathcal{Q}_s \right| = \left| \mathcal{Q}'_{q,m}\right| = \binom{q+3}{3} -4\binom{\frac{q}{2}+2}{3} $. 
Moreover, the $i$-th share has alphabet size $\left| \mathcal{Q}_{y_i} \right| =  \left| \mathcal{Q}_{g_i q,m} \right|= \binom{g_i q+3}{3}$. Equation \eqref{eq:sec} holds because the ratio of $\log_2 \left| \mathcal{Q}_{y_i} \right|$ and $\log_2 \left| \mathcal{Q}_{s} \right|$ approaches $1$ as $q \to \infty$.

For finite $q$ and $m=2$, the lower bound of $H(S|Y'')$ for Construction \ref{cnstr:RSSS} can be calculated from \eqref{eq:exact_lower} of Lemma \ref{lem: secrecy}. 
The difference between the lower bound \eqref{eq:exact_lower} and the upper bound \eqref{eq:exact_upper} is a constant dependent only on the matrix $G$ but not on the resolution $q$, as summarized in the following corollary.

\begin{corollary}\label{cor: secrecy_bounds}
Assume $m=2$ and consider an optimal construction (in terms of secrecy) satisfying the recovery condition \eqref{def: rec}, whose secret alphabet and share resolution are the same as Construction \ref{cnstr:RSSS}. For any $k-L+j$ shares of $Y''$ of $Y$, $j \in [0,L]$, the difference of $H(S|Y'')$ between the optimal construction and Construction \ref{cnstr:RSSS} is at most
\begin{align}\label{eq: leakage}
  \max_{\mathcal{I} \subseteq [n], |\mathcal{I}| = k} 
\quad \sum_{i \in \mathcal{I}} \log_2 g_i .
\end{align}
\end{corollary}
\begin{IEEEproof}
    We use the upper bound Equation \eqref{eq:exact_upper} and the lower bound Equation \eqref{eq:exact_lower}. Notice that the term $ \sum_{i=k-L+j+1}^{k}\log_2 g_i$ in \eqref{eq:exact_upper} corresponds to all shares in $Y_1^k$ but not in $Y''$, and the term $-\log_2 g=-\log_2\sum_{i=1}^{k-L+j} g_i$ in \eqref{eq:exact_lower} corresponds to all shares in $Y''$. The difference of the two terms simply corresponds to all shares in $Y_1^k$. Finally, note that $Y_1^k$ can be replaced by any other $k$ shares, and the corollary is proved.
\end{IEEEproof}

\begin{remark}
For a finite resolution $q$, the mixture of multiple vessels results in a small fraction of undesirable information leakage, as seen in Tables \ref{tab: comparison} and \ref{tab: comparison2}. While this section aims to prove that this undesirable leakage vanishes asymptotically, the methods of wire-tap channel II \cite{wiretapII, V.K., LDPCwire, Errorwire} can serve as a complementary strategy at the cost of a higher complexity: it can eliminate undesirable secret leakage, but requires the encoding (such as random binning) of the information into composite DNA shares at a {reduced code rate}. Moreover, the methods of wire-tap channel II still require generator matrices that result in low resolutions as constructed in Section \ref{sec: const}. 
\end{remark}

\section{Construction of generator matrices}\label{sec: const}
First, we propose two desirable properties of $G$. Second, we prove the existence of the generator matrices and show an upper bound on the required integer size. Finally, we present several constructions including array-code-based constructions, and evaluate them based on the two desirable properties. 

For the sake of clarity, we continue to use scalar probabilities throughout this section. Probability vectors will follow similar argument.
We further assume that the generator matrix $G$ contains only non-negative integers.

Recall that our generator matrix must satisfy two rank conditions:  (i) any $k$ block rows of $G$ are full rank, and (ii) any $k-L$ block rows and the last $k-L$ block columns should be full rank. Below we introduce two additional desirable properties of $G$: the small operational complexity and/or the small information leakage.

The complexity involved in reading and writing symbols is dependent on the resolution.
We assume that the resolution of the input secret is fixed. The resolution of the shares depends on the structure of the generator matrix, as shown in Equation \eqref{eq: res}.  
We define the \emph{operational complexity indicator} as the ratio between the resolution of the $i$-th share ($q_{y_i} = g_i q$) and the secret ($q$), maximized over all shares:
 \begin{align}\label{eq: OC}
     OC= \max_{i \in [n]} \left(\frac{q_{y_i}}{q}\right) = \max_{i \in [n]} g_i,
 \end{align}
where $g_i$ is the sum of the absolute values in $i$-th row of the generator matrix $G$.
Motivated by Corollary \ref{cor: secrecy_bounds}, for a given generator matrix $G$ in Construction \ref{cnstr:RSSS}, we define the \emph{information leakage indicator} as the product of all the row sums,
\begin{align}\label{eq: IL}
    IL = \prod_{i=1}^{n} g_i.    
\end{align}

\subsection{Existence of generator matrices}\label{sec: exist}
First, we prove that the generator matrix 
$G$ satisfying the rank conditions of Construction \ref{cnstr:RSSS} always exists. By tailoring the Combinatorial Nullstellensatz \cite{alon} to the set of integers, we can establish the existence of such a generator matrix. We then provide a small example derived from this existence proof. 

\begin{theorem}\label{exsitence}
    When $a> \binom{n}{k} +\binom{n}{k-L}$, there is an $n \times k$ generator matrix $G=(g_{i,j}), i\in[n], j\in[k],$ with integer elements $g_{i,j} \in [a]$ that satisfies the rank conditions in Construction \ref{cnstr:RSSS}.
\end{theorem}
\begin{IEEEproof}
    Consider a polynomial 
$f (g_{1,1}, g_{1,2}, \dots, g_{n,k})$ 
that represents the product of the determinants of all $k \times k$ submatrices and all submatrices consisting of $k-L$ rows and the last $k-L$ columns. There are $\binom{n}{k} + \binom{n}{k-L}$ determinant products in total. Let $t_{i,j}$ denote the exponent of each $g_{i,j}$, and the degree of $f$ is given by 
$\deg(f) = \sum_{i=1}^n \sum_{j=1}^k t_{i,j}$. 
Also, assume that the coefficient of $\prod_{i=1}^n \prod_{j=1}^k g_{i,j}$ in $f$ is non-zero.

Since each element in the matrix is used only once in the determinant calculation,  the maximum degree of $g_{i,j}$ is $t_{i,j}=\binom{n}{k} + \binom{n}{k-L}$. By applying Combinatorial Nullstellensatz II \cite{alon} and adapting it to polynomials over an integer set, we ensure that Theorem 1 of \cite{alon} and the assumptions in Theorem 2 of \cite{alon} remain valid. This adaptation confirms that working within an integer setting does not affect the applicability of Combinatorial Nullstellensatz II.

Thus, for all subsets $A_{i,j} \subseteq [a]$ such that $|A_{i,j}| > t_{i,j}$, there exist elements $g_{1,1} \in A_{1,1}, g_{1,2} \in A_{1,2}, \dots, g_{n,k} \in A_{n,k}$ such that 
$f  (g_{1,1}, g_{1,2}, \dots, g_{n,k}) \neq 0$. 
Therefore, when $a > \binom{n}{k} + \binom{n}{k-L}$, the generator matrix $G$ with these $g_{i,j}$ elements satisfies the rank conditions in Construction \ref{cnstr:RSSS}.
\end{IEEEproof}

\subsection{Construction and analysis}\label{sec: constex}

In the following, we provide some constructions of generator matrix $G$ and analyze the operational complexity  and information leakage. Numerical examples are provided at the end of this section. 
 
{\bf Random matrix.}
We uniformly at random choose a generator matrix with non-negative integer entries in the range $0$ to $\binom{n}{k} + \binom{n}{k-L}$. Based on Theorem \ref{exsitence}, there exists such a matrix satisfying the rank conditions. The operational complexity indicator is in the order of $k\left(\binom{n}{k} + \binom{n}{k-L}\right)$, and the information leakage indicator is in the order of $k^n\left(\binom{n}{k} + \binom{n}{k-L}\right)^n$. 

{\bf Vandermonde matrix\cite{vander1,vander2}.} By the properties of the Vandermonde matrix, it can meet our rank conditions. The Vandermonde matrix associated with the vector $[a_1, \ldots, a_n]$ is given by:
\begin{align}
V = \begin{bmatrix}
1 & a_1 & a_1^2 & \ldots & a_1^{k-1} \\
1 & a_2 & a_2^2 & \ldots & a_2^{k-1} \\
\vdots & \vdots & \vdots & \ddots & \vdots \\
1 & a_n & a_n^2 & \ldots & a_n^{k-1} \\
\end{bmatrix},
\end{align}
where  $\{a_i, i \in [n]\}$ are distinct non-zero values. 

Specifically, if we construct the Vandermonde matrix over a finite field $\mathbb{F}_N$, where $N > n$ and $N$ is a prime number,
the following properties hold over the field: (i) any $k$ block rows of $G$ are full rank, and (ii) any $k - L$ block rows along with the last $k - L$ block columns are also full rank.

Since the determinant of any such $k$ block rows of $G$ is non-zero over the finite field $\mathbb{F}_N$, it is also non-zero over the integers. The same argument applies to the rank condition (ii). Therefore, we can construct the Vandermonde matrix by selecting the $n$ smallest distinct positive integers for $[a_1, \dots, a_n]$. The powers of the vector elements can then be computed modulo $N$.
The operational complexity indicator can be bounded as
\begin{align}\label{eq: vandermonde1}
   OC= \max_{i \in [n]} g_i = \max_{i \in [n]}\sum_{j=1}^k g_{i,j} \le kN,
\end{align}
where $N$ is the smallest prime number such that $N > n$.
The information leakage indicator is bounded as
\begin{align}\label{eq: vandermonde2}
   IL = \prod_{i=1}^{n} g_i = \prod_{i=1}^n \sum_{j=1}^k g_{i,j} \le (kN)^n.
\end{align}

{\bf Cauchy matrix\cite{cauchy1840exercices}.} Consider the Cauchy matrix $G=(g_{i,j}), i\in[n], j\in[k]$, $g_{i,j}= \frac{1}{x_i-y_j}$, where $\{x_i= i \in [n]\}$, $\{y_j, j \in [k]\}$ are distinct. 
We choose $\{x_i= i \in [n]\}$, $\{y_j, j \in [k]\}$ to be distinct elements over a finite field $\mathbb{F}_N$, where $N\ge n+k$ is a prime number, and calculate all entries in $G$ over $\mathbb{F}_N$. Similar to the argument for the Vandermonde matrix, the resulting Cauchy matrix over the integers satisfies the rank conditions.    
The operational complexity indicator and the information leakage indicator can be bounded in the same way as \eqref{eq: vandermonde1} and \eqref{eq: vandermonde2}, except that $N$ is the smallest prime number such that $N \ge n+k$.

Here, we provide some numerical examples for the above generators.  Assume $n=5,k=3,L=1$. The rank conditions are (i) any $3\times 3$ submatrix is full rank $3$; (ii) any submatrix with $2$ rows and the last $2$ columns is full rank. 

\begin{example}\label{example:matrices}
     The Vandermonde matrix $G$ can be generated over the finite field $\mathbb{F}_{7}$, 
    \begin{align}
G = \begin{bmatrix}
1 & 1 & 1 \\
1 & 2 & 4 \\
1 & 3 & 2 \\
1 & 4 & 2 \\
1 & 5 & 4 \\
\end{bmatrix}.
    \end{align} 
The operational complexity indicator is $10$ and the information leakage indicator is $8830$. 

     The Cauchy matrix $G$ can be generated over the finite field $\mathbb{F}_{11}$, 
    \begin{align}
G = \begin{bmatrix}
    2 &   9  &  3 \\
    8 &   2  &  9 \\
    7 &   8  &  2 \\
    5 &   7  &  8 \\
   10 &   5  &  7 \\
\end{bmatrix}.
    \end{align}
The operational complexity indicator is $22$ and the information leakage indicator is $1.99 \times 10^6$. 

    Randomly select elements from $[0,20]$ to generate a matrix $G$, 
    \begin{align}
G = \begin{bmatrix}
18 & 9 & 10 \\
15 & 8 & 9 \\
6 & 16 & 14 \\
20 & 17 & 15 \\
0 & 4 & 16 \\
\end{bmatrix}.
    \end{align}
The two rank conditions can be verified for this example.
The operational complexity indicator is $52$ and the information leakage indicator is $4.43 \times 10^{7}$.
\end{example}

From the examples above, the Vandermonde generator has the lowest operational complexity indicator and information leakage indicator.  

After introducing the general generator for ARSSS, we also provide a special generator for $(k,1,k)$ ARSSS. In this special case, the secret size $L=1$, and the number of shares $n$ equals the number of shares $k$ required to recover the secret.

{\bf Modified circulant matrix.} For $(k,1,k)$ ARSSS, the generator matrix is of size $k \times k$. 
It is adapted from a circulant matrix with $2$ ones and $k-2$ zeros where each row (except the last row) is a cyclic shift of the previous row, and the last row has only a single one.
\begin{align}
G = 
\begin{bmatrix}
1 & 1 & 0 & \cdots & 0 \\
0 & 1 & 1 & \cdots & 0 \\
\vdots & \vdots & \vdots & \vdots & \vdots \\
0 & \cdots & \cdots & 0 & 1 \\
\end{bmatrix}.
\end{align}
This generator is full rank and any $k-1$ row with last $k-1$ column is full rank. It is obvious that the operational complexity indicator is $2$ and the information leakage indicator is $2^{k-1}$.

\subsection{Construction of generator matrices with array codes}\label{sec: array}
In the following, we investigate the scenario with a large amount of secret such that the size of a share exceeds the data capacity of a single vessel. Denote by $l$ the number of vessels to store one share.  
We consider these $l$ vessels collectively as a symbol, represented by a one-dimensional array of length $l$. Accordingly, encoding/decoding can be implemented by array codes, where each entry of the generator matrix is replaced by a sub-block matrix of size $l\times l$. 
Similar to Remark \ref{rmk: retrieval}, we can significantly reduce sequencing costs using the mixture operation even though the size of a share is increased to $l$ vessels.

In the following, we formally present the coding scheme, construct several generator matrices by leveraging known array codes, and demonstrate the advantage of array codes in terms of the operational complexity indicator and the information leakage indicator. 

\begin{construction}[Ramp secret sharing with array coding]
    Fix $n \ge k \ge L$.
Let the secret be $S=(s_1,\dots, s_L)^{T}$, where each symbol $s_i \in \mathbb{Z}_q^l, i \in [L]$ is an array with length $l$, and each element in the array is an integer uniformly selected from $[0,q]$. We generate the auxiliary sequence $X = (x_1,\dots,x_k)^{T}$, such that  $x_i=s_i, i\in [0, L]$ and $x_j, j\in [L+1, k]$ is an array uniformly selected from $[0,q]^l$. 
Let $G$ be an integer matrix with dimension $nl \times kl$. It can also be considered as the $n \times k$ block matrices, each block size is $l\times l$. Accordingly, we term $l$ rows (or columns) in a block to be a \emph{block row} (or a \emph{block column}). 
The rank conditions of generator $G$ are similar to Construction \ref{cnstr:RSSS}. The difference is to change the original rows (or columns) of $G$ to block rows (or block columns). 
The shares are generated from $Y=GX$, where each block row in $Y$ corresponds to one share.
\end{construction}

All proofs on the recovery and secrecy conditions will remain the same after changing the alphabet, and hence the above scheme is an ARSSS. Moreover, the operational complexity indicator and the information leakage indicator are defined in the same way as \eqref{eq: OC} and \eqref{eq: IL}, except that the matrix $G$ is of size $nl \times kl$ and $g_i =\sum_{j=1}^{kl}g_{i,j}$ denotes the sum of the $i$-th row in the matrix, for $i \in [nl]$.

{\bf A simple scheme.}
There is a simple construction of array generator which is achieved by multiplying each element in the non-array code generator of Construction \ref{cnstr:RSSS} with an identical matrix with size $l \times l$. This approach aligns with Theorem \ref{thm: 2} for probability vector when the array length $m=4$. Even though the two approaches coincide in terms of $G$, they differ in their physical interpretation: the array length $l$ represents the number of vessels per share, whereas the probability vector length $m$ corresponds to the number of different outcomes in the composite DNA letter. In constructing array codes, we need to consider block rows corresponding to the same array. It is straightforward to prove that if the non-array code generator satisfies the rank conditions, the array code construction will also fulfill the rank requirements. The operational complexity indicator remains the same as that of the non-array code generator, and the information leakage indicator can be simply extended from the non-array code generator.

The previous construction has a limitation in that all arrays must perform the same operations. In the following, we explore the more general array generators. 
To reduce operational complexity, we consider  binary array codes, and demonstrate two examples based on array codes in \cite{EVENODD, array}, while any other array code may also be adapted to our setting.

{\bf Schemes based on EVENODD codes.}
Consider the EVENODD array codes \cite{EVENODD}. They have codeword length $n'$, dimension $n'-2$, and array length $l=p-1$, where $p$ is a prime number and $n'\le p+2$. Consequently, we can construct ARSSS with parameters $(k=n'-2,L\le 2,n=n'-L)$.

The ideas are illustrated through the example code with $p=3,n'=4$, and the corresponding $(k=2,L=1, n=3)$ ARSSS. 
Assume our auxiliary sequence is
$X=[x_{1,1}, x_{2,1}, x_{1,2}, x_{2,2}]^{T}$. The first two elements are our secret and the last two elements are randomness. The EVENODD code has the $8\times 4$ generator matrix 
\begin{align}
     G = \begin{bmatrix}
1 & 0 & 0 & 0\\
0 & 1 &  0 &  0\\
\hline
0 & 0 &  1 &  0\\
0 & 0 &  0 &  1\\
1 & 0 &  1 &  0\\
0 & 1 &  0 &  1\\
1 & 0 &  0 &  1\\
0 & 1 &  1 &  1\\
     \end{bmatrix}.
\end{align}
Four shares can be constructed as $Y =GX= [y_{1,1}, y_{2,1}, y_{1,2}, y_{2,2} , y_{1,3}, y_{2,3}, y_{1,4}, y_{2,4}]^{T}$, where every two entries is one share. Since the first share ($y_{1,1},y_{2,1}$) is the secret, we discard it and only keep the other three shares $(y_{1,i}, y_{2,i}), i = 2,3, 4$. 
Correspondingly, the first two rows of $G$ are discarded.
Given that each array consists of two entries, the block rows/columns contain two consecutive rows/columns from $G$, highlighting a key distinction between array codes and non-array codes. Let $G_i, i=2,3,4$ represent the block row composed of the $(2i-1, 2i)$ rows in $G$. It is clear that any two of $G_i, i=2,3,4$ can form a $4\times4$ full rank matrix over the binary field, which means that any two shares can recover the secret. The last $2$ columns of $G_i$ is also a $2 \times 2$ full rank matrix over the binary field, for $i=2,3,4$, which means that any one share could not get any secret. 
Additionally, notice the following critical property: the determinant being one in the binary field guarantees that the determinant is an odd value over real numbers. Consequently, this generator $G$ without the first two rows meets our rank conditions. This ensures that the recovery and secrecy properties are preserved. The operational complexity indicator equals $3$ and information leakage indicator equals $24$.

{\bf Schemes based on polynomial ring.}
We also consider another array code based on polynomial ring \cite{array}. The array length is $l=p-1$, the codeword length is $n$, and the elements in the array are in $\mathbb{F}_q$, where $q$ is a prime power, $\text{gcd}(p,q)=1$, and $n \le p$. We will construct $(n,k,L)$ ARSSS for any $n \ge k \ge L$. These array codes have an efficient decoding algorithm which avoids multiplications over the extension field, instead utilizing simple cyclic shifts of vectors over $\mathbb{F}_q$. In particular, $q=2$ results in a binary generator matrix. 

We first describe the code using a non-array code format.
Let $\mathcal{R}_p(q)$ denote the ring of polynomials of degree less than $p-1$ over $\mathbb{F}_q$ with multiplication taken modulo $M_p(\alpha)= \sum_{i=0}^{p-1}\alpha^i$. The code over the alphabet $\mathcal{R}_p(q)$ 
is defined by the $(n-k) \times n$ parity-check matrix,
\begin{align}
    H' =  \begin{bmatrix}
1 & 1 & 1 & \dots & 1\\
1 & \alpha & \alpha^2 & \dots & \alpha^{n-1}\\
\vdots & \vdots & \vdots & \vdots &\vdots \\
1 & \alpha^{(n-k-1)} & \alpha^{2(n-k-1)} & \dots  & \alpha^{(n-k-1)(n-1)}
\end{bmatrix},
\end{align}
where $\alpha$ is the degree-1 monomial in $\mathcal{R}_p(q)$. 
Aiming at reducing the operational complexity and the information leakage, we modify it to the parity check matrix of a generalized Reed-Solomon code:
\begin{align}
    H =  H'T,
\end{align}
where $T$ is an $n \times n$ diagonal matrix
\begin{align}
    T =
    \begin{bmatrix}
t_1 & 0 & \dots & 0\\
0 & t_2  & \dots & 0\\
\vdots  & \vdots & \ddots &\vdots \\
0 & 0  & \dots  & t_{n}\\
\end{bmatrix},
\end{align}
for $t_i^{-1}= \prod_{j \in [n]\backslash \{i\}}(\alpha^{i} - \alpha^j), i\in[n]$.
Using the properties of generalized Reed-Solomon codes in Proposition 5.2 of \cite{Roth}, it can be verified that the corresponding $n\times k$ generator matrix is,
\begin{align}\label{eq:vanG}
    G = 
    \begin{bmatrix}
 1 & 1 & \dots & 1\\
1 & \alpha  & \dots & \alpha^{k-1}\\
\vdots  & \vdots & \vdots &\vdots \\
1 & \alpha^{(n-1)}  & \dots  & \alpha^{(k-1)(n-1)}\\
\end{bmatrix}.
\end{align}

Although the results presented in \cite{Roth} are for finite fields, they can also be extended to our polynomial ring. In particular, we need to show that $t_i$ is well defined. Knowing $\alpha^{p}=1$ from \cite{array}, $a^{-j}$ is valid and equal to $\alpha^{p-j}$. According to Lemma 1 in \cite{array}, the inverse of $\alpha^{i-j}-1$ in polynomial ring can be verified as $-\frac{1}{p}\sum_{w=0}^{p-2}(p-1-\alpha^w)$, where $\frac{1}{p}$ is also well-defined in $\mathbb{F}_q$. Consequently, the inverse of $\alpha^{j}(\alpha^{i-j}-1)$ is valid. Therefore, $t_i$ can be iteratively calculated. 

The rank conditions can be verified for the above non-array scheme. In particular, the corresponding submatrices are Vandermonde matrices over $\mathcal{R}_p(q)$ and are proved to be full rank in \cite{array}.

Furthermore, the above scheme can be represented by an array code. Here, we focus on the binary case with $q=2$. There exists a one-to-one mapping between the polynomial ring and the binary array when $q=2$. 
Assume $c$ is an element in $\mathcal{R}_p(2)$. Then $c$ can also be represented as $c = [c_{0},c_{1},\dots,c_{p-1}]^T \in \mathbb{F}_2^{p-1}$ such that
\begin{align}
    c = c(\alpha) =\sum_{i=0}^{p-2}c_{i}\alpha^i. 
\end{align}

Multiplying the element $\alpha^t$  with $c$ over $\mathcal{R}_p(2)$ can be illustrated as follows: 
\begin{align}
    \alpha^t \times c(\alpha) \mod{M_p(\alpha)}.
\end{align}
The multiplication in the polynomial ring is equivalent to cyclic shift and possible bit flip for the binary vector $[c_0,c_1,\dots,c_{p-1}]^T$ based on the facts that $\alpha^p = 1$ and $\alpha^{p-1} \equiv \sum_{i=0}^{p-2}\alpha^i \mod{M_p(\alpha)}$. Thus, we can  represent the factor $\alpha^t \in \mathcal{R}_p(2)$ by a $(p-1)\times (p-1)$ binary matrix $P$, where each element $P_{i, j}, i,j\in[0,p-2]$ is given by:
\begin{align}\label{eq:P_matrix}
P_{i, j} = 
\begin{cases}
1, & \text{if }   (j + t) \equiv i \text{ or  } p-1 \mod{p},  \\
0, & \text{otherwise.}
\end{cases}.
\end{align}
For example, when $p=5$, $c(\alpha)= c_{0}+ c_{1}\alpha + c_{2}\alpha^2 + c_{3}\alpha^3$. Then 
$\alpha^3 c(\alpha)$ is equivalent to

\begin{align}
    \begin{bmatrix}
0 & 1 & 1 & 0 \\
0 & 1 & 0 & 1 \\
0 & 1 & 0 & 0 \\
1 & 1 & 0 & 0
\end{bmatrix} \begin{bmatrix}
c_{0}  \\
c_{1} \\
c_{2}  \\
c_{3}
\end{bmatrix} = \begin{bmatrix}
c_{2}+c_{1}  \\
c_{3}+c_{1} \\
c_{1}  \\
c_{0}+ c_{1}
\end{bmatrix}.
\end{align}
Note that each row of $P$ has at most $2$ ones.

We replace each element of $G$ in \eqref{eq:vanG} by a binary block matrix in \eqref{eq:P_matrix}. 
The resulting generator matrix has at most $2k$ ones in each row, and thus the operational complexity indicator satisfies $OC \le 2k$, and the information leakage indicator satisfies $IL \le (2k)^{ln}$. The extra factor $l$ in the exponent of $IL$ is due to the larger alphabet of array codes, and if we calculate the ratio between information leakage and the secret, the factor $l$ will be eliminated. Therefore, this array scheme outperforms the non-array schemes in Section \ref{sec: constex}. Notably, the array schemes are suitable when the share size is larger than the vessel capacity.

Besides the above two classes of array codes, other array codes may also be modified to construct secret sharing schemes. For example, the systematic array code introduced in \cite{independent} offers efficient information updates, meaning the number of ones in the generator matrix is small. Thus, both operational complexity and information leakage can be reduced. 

Similar to EVENODD codes, to maintain rank condition (ii),
we need to truncate the systematic secret part from the array codeword and retain only the remaining part.

\section{Conclusion}\label{sec:conclusion}

This paper proposes an asymptotically ramp secret share scheme for probability vectors, motivated by composite DNA storage. Different from traditional ramp secret share schemes, probability vectors are not in a finite field. We develop a mixture operation between probability vectors, where subtraction and negative multiplication can be achieved.
An attractive property of the mixture operation herein is that it takes advantage of the original matrix products and efficiently achieves encoding and decoding, which dramatically reduces the synthesis cost and sequencing cost.

\bibliographystyle{IEEEtran}

\bibliography{bibligraphy.bib}

\begin{thebibliography}{10}
\providecommand{\url}[1]{#1}
\csname url@samestyle\endcsname
\providecommand{\newblock}{\relax}
\providecommand{\bibinfo}[2]{#2}
\providecommand{\BIBentrySTDinterwordspacing}{\spaceskip=0pt\relax}
\providecommand{\BIBentryALTinterwordstretchfactor}{4}
\providecommand{\BIBentryALTinterwordspacing}{\spaceskip=\fontdimen2\font plus
\BIBentryALTinterwordstretchfactor\fontdimen3\font minus \fontdimen4\font\relax}
\providecommand{\BIBforeignlanguage}[2]{{%
\expandafter\ifx\csname l@#1\endcsname\relax
\typeout{** WARNING: IEEEtran.bst: No hyphenation pattern has been}%
\typeout{** loaded for the language `#1'. Using the pattern for}%
\typeout{** the default language instead.}%
\else
\language=\csname l@#1\endcsname
\fi
#2}}
\providecommand{\BIBdecl}{\relax}
\BIBdecl

\bibitem{DNA1}
J.~Cox, ``Long-term data storage in {DNA},'' \emph{Trends in biotechnology}, vol.~19, pp. 247--50, 08 2001.

\bibitem{DNA2}
V.~Zhirnov, R.~Zadegan, G.~Sandhu, G.~Church, and W.~Hughes, ``Nucleic acid memory,'' \emph{Nature Materials}, vol.~15, pp. 366--370, 03 2016.

\bibitem{DNA3}
S.~M.~H. Tabatabaei~Yazdi, Y.~Yuan, J.~Ma, H.~Zhao, and O.~Milenkovic, ``A rewritable, random-access {DNA}-based storage system,'' \emph{Scientific Reports}, vol.~5, 05 2015.

\bibitem{DNA4}
H.~Lee, R.~Kalhor, N.~Goela, J.~Bolot, and G.~Church, ``Terminator-free template-independent enzymatic {DNA} synthesis for digital information storage,'' \emph{Nature Communications}, vol.~10, 06 2019.

\bibitem{DNA5}
G.~Church, Y.~Gao, and S.~Kosuri, ``Next-generation digital information storage in {DNA},'' \emph{Science (New York, N.Y.)}, vol. 337, p. 1628, 08 2012.

\bibitem{DNA6}
N.~Goldman, P.~Bertone, S.~Chen, C.~Dessimoz, E.~Leproust, B.~Sipos, and E.~Birney, ``Towards practical, high-capacity, low-maintenance information storage in synthesized {DNA},'' \emph{Nature}, vol. 494, 01 2013.

\bibitem{DNA7}
Y.~Erlich and D.~Zielinski, ``{DNA} fountain enables a robust and efficient storage architecture,'' \emph{Science}, vol. 355, pp. 950--954, 03 2017.

\bibitem{Anavy_DNA_letters}
L.~Anavy, I.~Vaknin, O.~Atar, R.~Amit, and Z.~Yakhini, ``Data storage in {DNA} with fewer synthesis cycles using composite {DNA} letters,'' \emph{Nature Biotechnology}, vol.~37, 10 2019.

\bibitem{choi}
Y.~Choi, T.~Ryu, A.~C. Lee, H.~Choi, H.~Lee, J.~Park, S.-H. Song, S.~Kim, H.~Kim, W.~Park \emph{et~al.}, ``High information capacity {DNA}-based data storage with augmented encoding characters using degenerate bases,'' \emph{Scientific reports}, vol.~9, no.~1, p. 6582, 2019.

\bibitem{molecular}
I.~Vaknin and R.~Amit, ``Molecular and experimental tools to design synthetic enhancers,'' \emph{Current Opinion in Biotechnology}, vol.~76, p. 102728, 2022.

\bibitem{data}
I.~Preuss, Z.~Yakhini, and L.~Anavy, ``Data storage based on combinatorial synthesis of dna shortmers,'' \emph{bioRxiv}, pp. 2021--08, 2021.

\bibitem{yan2023}
Y.~Yan, N.~Pinnamaneni, S.~Chalapati, C.~Crosbie, and R.~Appuswamy, ``Scaling logical density of {DNA} storage with enzymatically-ligated composite motifs,'' \emph{bioRxiv}, pp. 2023--02, 2023.

\bibitem{wenkai_full}
W.~Zhang and Z.~Wang, ``Secret sharing for dna probability vectors,'' in \emph{ICC 2024 - IEEE International Conference on Communications}, 2024, pp. 4578--4583.

\bibitem{sabary2024survey}
O.~Sabary, H.~M. Kiah, P.~H. Siegel, and E.~Yaakobi, ``Survey for a decade of coding for dna storage,'' \emph{IEEE Transactions on Molecular, Biological, and Multi-Scale Communications}, 2024.

\bibitem{icc2022}
W.~Zhang, Z.~Chen, and Z.~Wang, ``Limited-magnitude error correction for probability vectors in {DNA} storage,'' in \emph{ICC 2022 - IEEE International Conference on Communications}, 2022, pp. 3460--3465.

\bibitem{kobovich2023m}
A.~Kobovich, E.~Yaakobi, and N.~Weinberger, ``M-dab: An input-distribution optimization algorithm for composite {DNA} storage by the multinomial channel,'' \emph{arXiv preprint arXiv:2309.17193}, 2023.

\bibitem{zhang2024codes}
W.~Zhang and Z.~Wang, ``Codes for limited-magnitude probability error in dna storage,'' \emph{arXiv preprint arXiv:2405.10447}, 2024.

\bibitem{walter2024coding}
F.~Walter, O.~Sabary, A.~Wachter-Zeh, and E.~Yaakobi, ``Coding for composite dna to correct substitutions, strand losses, and deletions,'' \emph{arXiv preprint arXiv:2404.12868}, 2024.

\bibitem{cohen2024optimizing}
T.~Cohen and E.~Yaakobi, ``Optimizing the decoding probability and coverage ratio of composite dna,'' in \emph{2024 IEEE International Symposium on Information Theory (ISIT)}.\hskip 1em plus 0.5em minus 0.4em\relax IEEE, 2024, pp. 1949--1954.

\bibitem{sabary2024error}
O.~Sabary, I.~Preuss, R.~Gabrys, Z.~Yakhini, L.~Anavy, and E.~Yaakobi, ``Error-correcting codes for combinatorial composite dna,'' in \emph{2024 IEEE International Symposium on Information Theory (ISIT)}.\hskip 1em plus 0.5em minus 0.4em\relax IEEE, 2024, pp. 109--114.

\bibitem{kobovich2025deepdive}
A.~Kobovich, E.~Yaakobi, and N.~Weinberger, ``Deepdive: Optimizing input-constrained distributions for composite dna storage via multinomial channel,'' \emph{arXiv preprint arXiv:2501.15172}, 2025.

\bibitem{pierobon}
M.~Pierobon and I.~F. Akyildiz, ``A physical end-to-end model for molecular communication in nanonetworks,'' \emph{IEEE Journal on Selected Areas in Communications}, vol.~28, no.~4, pp. 602--611, 2010.

\bibitem{Farsad}
N.~Farsad, H.~B. Yilmaz, A.~Eckford, C.-B. Chae, and W.~Guo, ``A comprehensive survey of recent advancements in molecular communication,'' \emph{IEEE Communications Surveys \& Tutorials}, vol.~18, no.~3, pp. 1887--1919, 2016.

\bibitem{nakano2013molecular}
T.~Nakano, \emph{Molecular communication}.\hskip 1em plus 0.5em minus 0.4em\relax Cambridge University Press, 2013.

\bibitem{secover}
M.~M. Alsaffar, M.~Hasan, G.~P. McStay, and M.~Sedky, ``Digital {DNA} lifecycle security and privacy: an overview,'' \emph{Briefings in Bioinformatics}, vol.~23, no.~2, p. bbab607, 2022.

\bibitem{secover1}
M.~Backes, P.~Berrang, M.~Humbert, X.~Shen, and V.~Wolf, ``Simulating the large-scale erosion of genomic privacy over time,'' \emph{IEEE/ACM transactions on computational biology and bioinformatics}, vol.~15, no.~5, pp. 1405--1412, 2018.

\bibitem{WinNT}
\BIBentryALTinterwordspacing
MarketsandMarkets. (2023) {DNA} data storage companies - {Illumina, Inc. (US) and Microsoft (US)} are the key player. [Online]. Available: \url{https://www.marketsandmarkets.com/ResearchInsight/dna-data-storage-market.asp}
\BIBentrySTDinterwordspacing

\bibitem{sec0}
G.~R. BLAKLEY, ``Safeguarding cryptographic keys,'' in \emph{1979 International Workshop on Managing Requirements Knowledge (MARK)}, 1979, pp. 313--318.

\bibitem{sec2}
A.~Shamir, ``How to share a secret,'' \emph{Communications of the ACM}, vol.~22, no.~11, pp. 612--613, 1979.

\bibitem{de2016new}
P.~Y. De~Silva, G.~U. Ganegoda \emph{et~al.}, ``New trends of digital data storage in dna,'' \emph{BioMed research international}, vol. 2016, 2016.

\bibitem{ney2017computer}
P.~Ney, K.~Koscher, L.~Organick, L.~Ceze, and T.~Kohno, ``Computer security, privacy, and $\{$DNA$\}$ sequencing: compromising computers with synthesized $\{$DNA$\}$, privacy leaks, and more,'' in \emph{26th USENIX Security Symposium (USENIX Security 17)}, 2017, pp. 765--779.

\bibitem{zhang2022preservation}
Y.~Zhang, Y.~Ren, Y.~Liu, F.~Wang, H.~Zhang, and K.~Liu, ``Preservation and encryption in dna digital data storage,'' \emph{ChemPlusChem}, vol.~87, no.~9, p. e202200183, 2022.

\bibitem{DNA_secret0}
C.~T. Clelland, V.~Risca, and C.~Bancroft, ``Hiding messages in {DNA} microdots,'' \emph{Nature}, vol. 399, no. 6736, pp. 533--534, 1999.

\bibitem{DNA_secret1}
A.~Gehani, T.~LaBean, and J.~Reif, ``{DNA}-based cryptography,'' in \emph{Aspects of Molecular Computing}.\hskip 1em plus 0.5em minus 0.4em\relax Springer, 2003, pp. 167--188.

\bibitem{DNA_secret2}
A.~Leier, C.~Richter, W.~Banzhaf, and H.~Rauhe, ``Cryptography with {DNA} binary strands,'' \emph{Biosystems}, vol.~57, no.~1, pp. 13--22, 2000.

\bibitem{2010grammar}
M.~R. Ogiela and U.~Ogiela, ``Grammar encoding in {DNA}-like secret sharing infrastructure,'' in \emph{International Conference on Advanced Computer Science and Information Technology}.\hskip 1em plus 0.5em minus 0.4em\relax Springer, 2010, pp. 175--182.

\bibitem{Adhikari}
A.~Adhikari, ``{DNA} secret sharing,'' in \emph{2006 IEEE International Conference on Evolutionary Computation}, 2006, pp. 1407--1411.

\bibitem{ramp1}
G.~Blakley and C.~Meadows, ``Security of ramp schemes,'' vol. 196, 01 1984, pp. 242--268.

\bibitem{ramp2}
\BIBentryALTinterwordspacing
R.~J. McEliece and D.~V. Sarwate, ``On sharing secrets and reed-solomon codes,'' \emph{Commun. ACM}, vol.~24, no.~9, p. 583–584, sep 1981. [Online]. Available: \url{https://doi.org/10.1145/358746.358762}
\BIBentrySTDinterwordspacing

\bibitem{ramp3}
P.~Paillier, ``On ideal non-perfect secret sharing schemes,'' in \emph{International Workshop on Security Protocols}.\hskip 1em plus 0.5em minus 0.4em\relax Springer, 1997, pp. 207--216.

\bibitem{ramp4}
A.~De~Santis and B.~Masucci, ``Multiple ramp schemes,'' \emph{IEEE Transactions on Information Theory}, vol.~45, no.~5, pp. 1720--1728, 1999.

\bibitem{yamamoto}
H.~Yamamoto, ``Secret sharing system using (k, l, n) threshold scheme,'' \emph{Electronics and Communications in Japan (Part I: Communications)}, vol.~69, no.~9, pp. 46--54, 1986.

\bibitem{MDS}
F.~J. MacWilliams and N.~J.~A. Sloane, \emph{The theory of error-correcting codes}.\hskip 1em plus 0.5em minus 0.4em\relax Elsevier, 1977, vol.~16.

\bibitem{pieprzyk}
J.~Pieprzyk and X.-M. Zhang, ``Ideal threshold schemes from mds codes,'' in \emph{Information Security and Cryptology—ICISC 2002: 5th International Conference Seoul, Korea, November 28--29, 2002 Revised Papers 5}.\hskip 1em plus 0.5em minus 0.4em\relax Springer, 2003, pp. 253--263.

\bibitem{chen2007}
H.~Chen, R.~Cramer, S.~Goldwasser, R.~De~Haan, and V.~Vaikuntanathan, ``Secure computation from random error correcting codes,'' in \emph{Annual International Conference on the Theory and Applications of Cryptographic Techniques}.\hskip 1em plus 0.5em minus 0.4em\relax Springer, 2007, pp. 291--310.

\bibitem{Umberto}
U.~Martínez-Peñas, ``Communication efficient and strongly secure secret sharing schemes based on algebraic geometry codes,'' \emph{IEEE Transactions on Information Theory}, vol.~64, no.~6, pp. 4191--4206, 2018.

\bibitem{wiretapII}
L.~H. Ozarow and A.~D. Wyner, ``Wire-tap channel {II},'' \emph{AT\&T Bell Laboratories Technical Journal}, vol.~63, no.~10, pp. 2135--2157, 1984.

\bibitem{V.K.}
V.~Wei, ``Generalized hamming weights for linear codes,'' \emph{IEEE Transactions on Information Theory}, vol.~37, no.~5, pp. 1412--1418, 1991.

\bibitem{LDPCwire}
A.~Thangaraj, S.~Dihidar, A.~R. Calderbank, S.~W. McLaughlin, and J.-M. Merolla, ``Applications of ldpc codes to the wiretap channel,'' \emph{IEEE Transactions on Information Theory}, vol.~53, no.~8, pp. 2933--2945, 2007.

\bibitem{Errorwire}
M.~Bloch, M.~Hayashi, and A.~Thangaraj, ``Error-control coding for physical-layer secrecy,'' \emph{Proceedings of the IEEE}, vol. 103, no.~10, pp. 1725--1746, 2015.

\bibitem{mix0}
L.~M. Adleman, ``On constructing a molecular computer.'' \emph{{DNA} based computers}, vol.~27, pp. 1--21, 1995.

\bibitem{mix1}
M.~Amos, A.~Gibbons, and D.~A. Hodgson, ``Error-resistant implementation of {DNA} computations,'' 1996.

\bibitem{mix2}
S.~Roweis, E.~Winfree, R.~Burgoyne, N.~V. Chelyapov, M.~F. Goodman, P.~W. Rothemund, and L.~M. Adleman, ``A sticker-based model for {DNA} computation,'' \emph{Journal of Computational Biology}, vol.~5, no.~4, pp. 615--629, 1998.

\bibitem{choi2019}
Y.~Choi, T.~Ryu, A.~C. Lee, H.~Choi, H.~Lee, J.~Park, S.-H. Song, S.~Kim, H.~Kim, W.~Park \emph{et~al.}, ``High information capacity dna-based data storage with augmented encoding characters using degenerate bases,'' \emph{Scientific reports}, vol.~9, no.~1, p. 6582, 2019.

\bibitem{lu2022enzymatic}
X.~Lu, J.~Li, C.~Li, Q.~Lou, K.~Peng, B.~Cai, Y.~Liu, Y.~Yao, L.~Lu, Z.~Tian \emph{et~al.}, ``Enzymatic dna synthesis by engineering terminal deoxynucleotidyl transferase,'' \emph{ACS Catalysis}, vol.~12, no.~5, pp. 2988--2997, 2022.

\bibitem{composite_motifs}
Y.~Yan, N.~Pinnamaneni, S.~Chalapati, C.~Crosbie, and R.~Appuswamy, ``Scaling logical density of dna storage with enzymatically-ligated composite motifs,'' \emph{Scientific Reports}, vol.~13, no.~1, p. 15978, 2023.

\bibitem{shortmer}
I.~Preuss, M.~Rosenberg, Z.~Yakhini, and L.~Anavy, ``Efficient dna-based data storage using shortmer combinatorial encoding,'' \emph{Scientific reports}, vol.~14, no.~1, p. 7731, 2024.

\bibitem{Zippel1993}
R.~Zippel, \emph{Diophantine Equations}.\hskip 1em plus 0.5em minus 0.4em\relax Boston, MA: Springer US, 1993, pp. 41--55.

\bibitem{1969diophantine}
L.~J. Mordell, \emph{Diophantine equations}.\hskip 1em plus 0.5em minus 0.4em\relax Academic press, 1969.

\bibitem{2006diophantine}
W.~M. Schmidt, \emph{Diophantine approximations and Diophantine equations}.\hskip 1em plus 0.5em minus 0.4em\relax Springer, 2006.

\bibitem{2001solving}
A.~Bockmayr and V.~Weispfenning, ``Solving numerical constraints,'' 2001.

\bibitem{2010diophantine}
T.~Andreescu, D.~Andrica, I.~Cucurezeanu \emph{et~al.}, \emph{An introduction to Diophantine equations: A problem-based approach}.\hskip 1em plus 0.5em minus 0.4em\relax Springer, 2010.

\bibitem{alabdulmohsin2018summability}
I.~M. Alabdulmohsin, \emph{Summability calculus: A comprehensive theory of fractional finite sums}.\hskip 1em plus 0.5em minus 0.4em\relax Springer, 2018.

\bibitem{alon}
N.~Alon, ``Combinatorial nullstellensatz,'' \emph{Combinatorics, Probability and Computing}, vol.~8, no. 1-2, pp. 7--29, 1999.

\bibitem{vander1}
H.~Roger and R.~J. Charles, ``Topics in matrix analysis,'' 1994.

\bibitem{vander2}
G.~H. Golub and C.~F. Van~Loan, \emph{Matrix computations}.\hskip 1em plus 0.5em minus 0.4em\relax JHU press, 2013.

\bibitem{cauchy1840exercices}
A.~L. Cauchy, \emph{Exercices d'analyse et de physique mathematique: 1}.\hskip 1em plus 0.5em minus 0.4em\relax Bachelier, imprimeur-libraire, 1840, vol.~1.

\bibitem{EVENODD}
M.~Blaum, J.~Brady, J.~Bruck, and J.~Menon, ``Evenodd: an efficient scheme for tolerating double disk failures in raid architectures,'' \emph{IEEE Transactions on Computers}, vol.~44, no.~2, pp. 192--202, 1995.

\bibitem{array}
M.~Blaum and R.~Roth, ``New array codes for multiple phased burst correction,'' \emph{IEEE Transactions on Information Theory}, vol.~39, no.~1, pp. 66--77, 1993.

\bibitem{Roth}
R.~Roth, \emph{Introduction to Coding Theory}.\hskip 1em plus 0.5em minus 0.4em\relax Cambridge University Press, 2006.

\bibitem{independent}
M.~Blaum, J.~Bruck, and A.~Vardy, ``{MDS} array codes with independent parity symbols,'' \emph{IEEE Transactions on Information Theory}, vol.~42, no.~2, pp. 529--542, 1996.

\end{thebibliography}


\end{document}